\documentclass{birkjour}
 \theoremstyle{definition}
 
 \theoremstyle{remark}

 \numberwithin{equation}{section}
 
\usepackage{pstricks}
\usepackage{pstricks-add}
\usepackage{enumitem}

\newcommand{\dd}{\mathrm{d}}
\newcommand{\R}{\mathbb{R}}
\newcommand{\Exp}{\mathbb{E}}
\renewcommand{\Pr}{\mathbb{P}}

\newcommand{\dto}{\downarrow}
\newcommand{\expo}[1]{\mathrm{e}^{#1}}

\newcommand{\Hess}{\mathrm{Hess}\,}

\newcommand{\Diff}{\mathrm{D}}

\newcommand{\transp}{\top}

\newcommand{\Cs}{\mathrm{C}}
\newcommand{\ACs}{\mathrm{H}^1}
\newcommand{\Ls}{\mathrm{L}}

\newcommand{\scal}[2]{\langle #1, #2\rangle}
\newcommand{\Scal}[2]{\left\langle #1, #2\right\rangle}
\renewcommand{\div}{\mathrm{div}~}

\newcommand{\Cst}{C_{\mathrm{st}}}
\newcommand{\Cbl}{C_{\mathrm{bl}}}
\newcommand{\Pst}{P_{\mathrm{st}}}
\newcommand{\pst}{p_{\mathrm{st}}}

\newcommand{\pqst}{p_{\mathrm{qst}}}
\newcommand{\jqst}{j_{\mathrm{qst}}}
\newcommand{\lqst}{\lambda_{\mathrm{qst}}}
\newcommand{\pabs}{p_{\mathrm{abs}}}

\newcommand{\Hs}{{H_{\star}}}
\newcommand{\Ds}{{D_{\star}}}
\newcommand{\Ss}{{\sigma_{\star}}}
\newcommand{\as}{{a_{\star}}}
\newcommand{\Ms}{{M_{\star}}}
\newcommand{\Ns}{{N_{\star}}}
\newcommand{\ns}{{n_{\star}}}

\DeclareMathOperator*{\Sim}{\sim}
\DeclareMathOperator*{\Asymp}{\asymp}

\begin{document}

%
%
%
%
%
%
%
%
%

\title[Irreversible Eyring-Kramers formula]
 {Generalisation of the Eyring-Kramers\\ transition rate formula to irreversible\\ diffusion processes}

\author[Freddy Bouchet]{Freddy Bouchet}

\address{%
Laboratoire de Physique\\
\'Ecole Normale Sup\'erieure de Lyon\\
46 all\'ee d'Italie\\
F-69364 Lyon}

\email{freddy.bouchet@ens-lyon.fr}

\thanks{The research leading to these results has received funding from the European Research Council under the European Union's seventh Framework Programme (FP7/2007-2013 Grant Agreement no. 616811) (F. Bouchet, and J. Reygner).}
\author[Julien Reygner]{Julien Reygner}
\address{%
Laboratoire de Physique\\
\'Ecole Normale Sup\'erieure de Lyon\\
46 all\'ee d'Italie\\
F-69364 Lyon}
\email{julien.reygner@polytechnique.org}



\begin{abstract}
In the small noise regime, the average transition time between metastable states of a reversible diffusion process is described at the logarithmic scale by Arrhenius' law. The Eyring-Kramers formula classically provides a subexponential prefactor to this large deviation estimate. For  irreversible diffusion processes, the equivalent of Arrhenius' law is given by the Freidlin-Wentzell theory. In this paper, we compute the associated prefactor and thereby generalise the Eyring-Kramers formula to irreversible diffusion processes. In our formula, the role of the potential is played by Freidlin-Wentzell's quasipotential, and a correction depending on the non-Gibbsianness of the system along the instanton is highlighted. Our analysis relies on a WKB analysis of the quasistationary distribution of the process in metastable regions, and on a probabilistic study of the process in the neighbourhood of saddle-points of the quasipotential.
\end{abstract}

\maketitle


\section{Introduction}


\subsection{The Eyring-Kramers formula} Many equilibrium systems in statistical physics can be described by the overdamped diffusion of a particle in $\R^d$ according to the solution $(X^{\epsilon}_t)_{t \geq 0}$ of the stochastic differential equation
\begin{equation}
  \dd X^{\epsilon}_t = -\nabla U(X^{\epsilon}_t) \dd t + \sqrt{2\epsilon} \dd W_t
\end{equation}
in $\R^d$, where $(W_t)_{t \geq 0}$ is a $d$-dimensional Brownian motion, $\epsilon > 0$ is a temperature parameter and $U : \R^d \to \R$ is the {\em potential} of the system. Upon integrability assumptions on this potential, the process is reversible and ergodic with respect to the Gibbs measure
\begin{equation}\label{eq:Gibbs:intro}
  p^{\epsilon}_{\mathrm{st}}(x) = \frac{1}{Z^{\epsilon}} \exp\left(-\frac{U(x)}{\epsilon}\right), \qquad Z^{\epsilon} := \int_{x \in \R^d} \exp\left(-\frac{U(x)}{\epsilon}\right) \dd x,
\end{equation}
which in particular implies that the particle visits infinitely often every region of the state space $\R^d$. However, when $\epsilon$ is small and $U$ possesses several local minima, or potential wells, the particle typically remains stuck at the bottom of these wells over long times, and transitions between different wells are rare events of the system. This is the metastability phenomenon.

\begin{figure}[ht]
  \begin{pspicture}(6,3)
    \rput(3,1.5){\includegraphics[clip=true,trim=3cm 4.5cm 2.8cm 4cm,width=6cm]{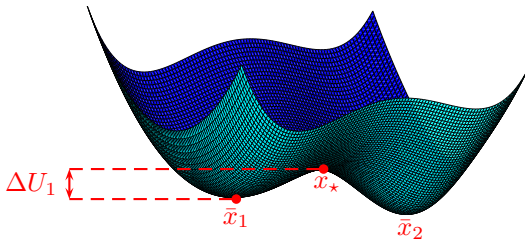}}
    \rput(2.2,.15){\red $\bar{x}_1$}
    \rput(3.4,.6){\red $x_{\star}$}
    \rput(4.5,0){\red $\bar{x}_2$}
    \psline[linecolor=red,linestyle=dashed]{-*}(0,.4)(2.2,.4)
    \psline[linecolor=red,linestyle=dashed]{-*}(0,.8)(3.35,.8)
    \psline[linecolor=red,linestyle=dashed]{<->}(0,.4)(0,.8)
    \rput(-.5,.6){\red$\Delta U_1$}
  \end{pspicture}
  \caption{Example of a double-well potential in dimension $d=2$. The potential barrier to escape from the well corresponding to the local minimum $\bar{x}_1$ is $\Delta U_1 = U(x_{\star})-U(\bar{x}_1)$.}
  \label{fig:doublewell}
\end{figure}

It is of course of major interest to quantify the time scale over which such transitions between different metastable states occur. A first answer is provided by Arrhenius' law~\cite{Arr89}, asserting that the average exit time $\Exp[\tau^{\epsilon}]$ from a potential well satisfies the logarithmic equivalence
\begin{equation}
  \lim_{\epsilon \dto 0} \epsilon \log \Exp[\tau^{\epsilon}] = \Delta U,
\end{equation}
where $\Delta U$ is the minimal potential barrier that the particle has to climb in order to escape from the well, see Figure~\ref{fig:doublewell}. In the sequel of the paper, we shall use the notation 
\begin{equation}\label{eq:arrh}
  \Exp[\tau^{\epsilon}] \Asymp_{\epsilon \dto 0} \exp\left(\frac{\Delta U}{\epsilon}\right)
\end{equation}
to refer to such a logarithmic equivalence.

Let two potential wells, corresponding to local minima $\bar{x}_1$ and $\bar{x}_2$, communicate through a saddle-point $x_{\star}$ (see Figure~\ref{fig:doublewell}). We make the generical assumption that the Hessian matrix $\Hess U(\bar{x}_1)$ is positive-definite, and that the Hessian matrix $\Hess U(x_{\star})$ has one negative eigenvalue $-\lambda^{\star}_+$, and $d-1$ positive eigenvalues. Let us denote by $\tau^{\epsilon}_{\bar{x}_1 \to \bar{x}_2}$ the transition time between the two wells. Then the subexponential correction to Arrhenius' law is given by the Eyring-Kramers formula
\begin{equation}\label{eq:EKrev}
  \Exp[\tau^{\epsilon}_{\bar{x}_1 \to \bar{x}_2}] \Sim_{\epsilon \dto 0} \frac{2\pi}{\lambda^{\star}_+} \sqrt{\frac{|\det \Hess U(x_{\star})|}{\det \Hess U(\bar{x}_1)}} \times \exp\left(\frac{\Delta U}{\epsilon}\right).
\end{equation}
The formula is named after Eyring and Kramers' respective papers~\cite{Eyr35,Kra40}. Although we do not exactly know when this formula was first derived for dynamics in any dimensions, it already appeared for that case in Landauer and Swanson~\cite{LanSwa61}, and Langer~\cite{Lan69}.

The mathematical proof of this formula, as well as its extension to more general metastable situations, was first obtained by Bovier, Eckhoff, Gayrard, and Klein~\cite{BovEckGayKle04} by means of a potential theoretic approach, and shortly after by Helffer, Klein and Nier~\cite{HelKleNie04} through Witten Laplacian analysis. We refer to the survey by Berglund~\cite{Ber13} for a review of mathematical approaches to the Eyring-Kramers formula, and highlight that the study of such formulas for general equilibrium systems remains an active field in mathematics, see for instance~\cite{BerGen10, BouTou12, BerDut13} for degenerate potentials, \cite{MenSch14} for the relation with functional inequalities, or~\cite{BarBovMel10, BerGen13, Bar15, DiGLeP15} for infinite-dimensional models described by stochastic partial differential equations and associated particle systems.


\subsection{Metastability out of equilibrium}\label{ss:intro:metaneq} The purpose of this article is to establish a similar Eyring-Kramers formula for general diffusion processes
\begin{equation}\label{eq:SDE:intro}
  \dd X^{\epsilon}_t = b(X^{\epsilon}_t) \dd t + \sqrt{2\epsilon}\sigma(X^{\epsilon}_t) \dd W_t,
\end{equation}
that are not necessarily reversible, so that they can describe physical systems out of equilibrium. In the small noise asymptotics, the particle typically fluctuates around the attractors of the zero-noise dynamics
\begin{equation}\label{eq:detdyn}
  \dot{x}_t = b(x_t),
\end{equation}
which are the metastable states of the system. In this paper, the zero-noise dynamics is called {\em relaxation dynamics}, and we shall assume that its only attractors are isolated points. We refer to~\cite{Gra88} or~\cite{BerGen04, BerGen14, Ber14} for a description of the behaviour of more general nonequilibrium systems in the small noise regime. In our simple situation, the metastability of the process is expressed by the fact that the particle spends most of its time fluctuating in the neighbourhood of attractors, where the law of its position is described by a {\em quasistationary} distribution, and rarely jumps from one attractor to another one, under the effect of the noise. The typical paths employed by the particle for such fluctuations far from the typical behaviour are called {\em fluctuation paths}.

At the logarithmic scale, the expected transition time between two attractors is described by the Freidlin-Wentzell theory~\cite{FreWen12}, which extends Arrhenius' law to irreversible processes. In this theory, the role of the potential in the reversible case is played by a quantity called {\em quasipotential}, constructed in terms of the action of fluctuation paths. In a simple situation where the relaxation dynamics~\eqref{eq:detdyn} possesses two attractors $\bar{x}_1$ and $\bar{x}_2$, the transition time from $\bar{x}_1$ to $\bar{x}_2$ generally satisfies
\begin{equation}\label{eq:FW:intro}
  \Exp[\tau^{\epsilon}_{\bar{x}_1 \to \bar{x}_2}] \Asymp_{\epsilon \dto 0} \exp\left(\frac{V(\bar{x}_1,x_{\star})}{\epsilon}\right),
\end{equation}
where $V(\bar{x}_1, \cdot)$ is the quasipotential with respect to $\bar{x}_1$ and $x_{\star}$ is a relevant saddle-point of the quasipotential, located on the hypersurface separating the respective basins of attraction of $\bar{x}_1$ and $\bar{x}_2$. When $\epsilon \dto 0$, the typical trajectory employed by the process to reach the saddle-point $x_{\star}$ is deterministic. It is referred to as the {\em instanton}, or sometimes {\em most probable escape path}~\cite{MaiSte97}, and is denoted by $(\rho_t)_{t \in \R}$. It connects $\bar{x}_1$ to $x_{\star}$ in infinite time, so that it satisfies
\begin{equation}
  \lim_{t \to -\infty} \rho_t = \bar{x}_1, \qquad \lim_{t \to +\infty} \rho_t = x_{\star}.
\end{equation}


\subsection{Main results and outline of the article} The purpose of this article is to supply~\eqref{eq:FW:intro} with a subexponential prefactor, similarly to the Eyring-Kramers formula. We mostly rely on two ingredients: a perturbative calculation of the quasistationary distribution in the neighbourhood of attractors, and a probabilistic study of the diffusion process~\eqref{eq:SDE:intro} in the neighbourhood of saddle-points. 

Let us explain our argument in further detail. In Section~\ref{s:qp}, we begin by recalling some notions from the Freidlin-Wentzell theory~\cite{FreWen12} of large deviations for the diffusion process~\eqref{eq:SDE:intro}. In particular, we give the definition of the quasipotential with respect to an attractor of the relaxation dynamics, and insist on the fact that this quasipotential can be identified when the vector field $b$ admits a decomposition into a gradient part and a transverse force. We then discuss the formulation of this transverse decomposition in terms of a Hamilton-Jacobi equation for the quasipotential, which is related with the regularity of the quasipotential and plays a crucial role in the sequel of the paper. The study of the regularity of the quasipotential is a nontrivial mathematical issue and leads to interesting physical phenomena~\cite{BaeKaf15}. Throughout the paper, we avoid considering such phenomena, and we shall rather establish our main result under the assumption that the quasipotential be smooth in the neighbourhood of the instanton. Assuming that the quasipotential be smooth in a neighborhood of the instanton is not so restrictive, as this situation is thought of being generic, even in situations when Lagrangian singularities do exists in some part of the phase space.

Sections~\ref{s:stat} and~\ref{s:bl} are dedicated to the study of the quasistationary distribution of the diffusion process~\eqref{eq:SDE:intro} in the neighbourhood of an attractor. In Section~\ref{s:stat}, we compute the prefactor to the {\em ensemble measure}, that is, the nonequilibrium stationary distribution of a particle diffusing in a vector field with a single global attractor. At the logarithmic scale, the Freidlin-Wentzell theory implies that the ensemble measure is equivalent to a Gibbs measure with the quasipotential playing the role of the potential. Using a classical WKB approach~\cite{MatSch77, SchMat79, Gra88, MaiSte93, MaiSte97}, we show that the transport equation satisfied by the prefactor to this logarithmic equivalent can be solved using the most probable fluctuation paths of the process as characteristics, or rays~\cite{CohLew67, Lud75, MatSch77}, and thereby obtain an explicit expression for this prefactor which provides a correction taking into account the non-Gibbsianness of the system along these fluctuation paths. In Section~\ref{s:bl}, we fix a domain $D \subset \R^d$ attracted to a single equilibrium point of the relaxation dynamics~\eqref{eq:detdyn}, and perform a boundary layer analysis of the quasistationary distribution in $D$ to compute the associated probability current at some point $y \in \partial D$. Integrating this current over $\partial D$ yields the rate $\lqst^{\epsilon}$ at which a particle escapes from $D$ under the quasistationary distribution.

In the bistable situation described in Subsection~\ref{ss:intro:metaneq}, the irreversible (or nonequilibrium) Eyring-Kramers formula is finally derived in Section~\ref{s:EK}. In contrast with Kramers' method~\cite{SchMat79, MaiSte93, MaiSte97, AriVdE07, Sch09}, we do not apply a WKB approximation to the elliptic problem solved by the average transition time, but rather combine the results of Section~\ref{s:bl} with a purely probabilistic study of the diffusion process in the neighbourhood of the saddle-point, in order to compute the rate at which a particle started from $\bar{x}_1$ reaches the neighbourhood of $\bar{x}_2$. Taking the inverse of this rate yields the following expression for the average transition time
\begin{equation}\label{eq:EKneq}
  \begin{aligned}
    \Exp[\tau^{\epsilon}_{\bar{x}_1 \to \bar{x}_2}] & \Sim_{\epsilon \dto 0} \frac{2\pi}{\lambda^{\star}_+} \sqrt{\frac{|\det H_{\star}|}{\det \Hess_x V(\bar{x}_1,\bar{x}_1)}} \exp\left(\int_{-\infty}^{+\infty} F(\rho_t)\dd t\right)\\
    & \quad \times \exp\left(\frac{V(\bar{x}_1,x)}{\epsilon}\right).
  \end{aligned}
\end{equation}
In this irreversible Eyring-Kramers formula, $\Hs$ has to be understood as the suitable definition of the Hessian of the quasipotential at the saddle-point, so that the ratio of Hessian determinants is a straightforward generalisation of the equilibrium case, with the quasipotential playing the role of the potential. Besides, $\lambda^{\star}_+$ is the positive eigenvalue of the Jacobian matrix $\Diff b(x_{\star})$, corresponding to the unstable direction of the vector field at the saddle-point. Note that, in general, $\lambda^{\star}_+$ is not an eigenvalue of $-\Hs$, so that it depends on the whole vector field $b$ and not only on its gradient part. The last correction with respect to the equilibrium Eyring-Kramers formula is the term involving the integral of the function $F$ along the instanton $(\rho_t)_{t \in \R}$. This function $F$ arises from the prefactor to the quasistationary distribution, and has to be interpreted as a measure of the non-Gibbsianness of the system, in the sense that $F \equiv 0$ if and only if the Gibbs measure~\eqref{eq:Gibbs:intro} is stationary for the process~\eqref{eq:SDE:intro}.

Since our analysis is local, we expect this formula to extend to more general metastable situations, with more than two isolated attractors. 

Let us precise that similar nonequilibrium formulas have already been derived, although as far as we know, our probabilistic approach is new and leads to the original formula~\eqref{eq:EKneq}. In specific restricted cases, Maier and Stein~\cite{MaiSte93, MaiSte97} obtained a two-dimensional formula, while Ariel and Vanden-Eijnden~\cite{AriVdE07} addressed the case of irreversible diffusion processes preserving the invariance of the Gibbs measure. Based on a purely WKB analysis, Schuss~\cite{Sch09} obtained a preliminary version of~\eqref{eq:EKneq} that does not explicitly involve the Hessian matrix $\Hs$ of the quasipotential at the saddle-point. Besides, it seems to us that in none of these results, the nonequilibrium correction arising from the quasistationary behaviour was clearly identified in terms of the function $F$ and of the instanton.

We finally underline that our analysis relies on the assumptions of local smoothness of the quasipotential around the instanton, and of nondegeneracy of the Hessian matrix of the quasipotential at the saddle-point. We believe that these assumptions are generic. For instance, for generic vector fields $b$, Lagrangian singularities of the quasipotential do not occur along the instanton. These assumptions essentially lead to the existence of a transverse decomposition for the vector field $b$, which somehow allows us to combine WKB approximations with notions from the theory of large deviations. In particular, the most probable fluctuation paths of the diffusion process are rather expressed in terms of this transverse decomposition than as the solution to the Hamilton equations in a $2d$-dimensional phase space, which is the point of view of most previous papers~\cite{DykSmeMaiSil96,MaiSte97} where singularities of the quasipotential are addressed. As we shall discuss below, our assumptions in particular exclude some specific phenomena, such as the {\em grazing instantons} observed in~\cite{DykSmeMaiSil96,MaiSte97}.

\subsection{Notations} Throughout this paper, the scalar product in $\R^d$ is denoted by $\scal{\cdot}{\cdot}$. Given a function $f : \R^d \to \R$, we write $\partial_i f := \frac{\partial}{\partial x_i} f$ and $\partial_{ij} f := \frac{\partial^2}{\partial x_i\partial x_j} f$. The Hessian matrix $\Hess f(x)$ is defined by $(\Hess f(x))_{ij} := \partial_{ij} f(x)$. Given a vector field $h = (h_1, \ldots, h_d) : \R^d \to \R^d$, the Jacobian matrix $\Diff h(x)$ is defined by $(\Diff h(x))_{ij} := \partial_j f_i(x)$.


\section{Quasipotential and transverse decomposition}\label{s:qp}

In this section, we introduce a few notions that are related with the Freidlin-Wentzell theory of the stochastic differential equation
\begin{equation}\label{eq:SDE}
  \dd X^{\epsilon}_t = b(X^{\epsilon}_t) \dd t + \sqrt{2\epsilon} \sigma(X^{\epsilon}_t)\dd W_t
\end{equation}
in $\R^d$, where $b : \R^d \to \R^d$ and $\sigma : \R^d \to \R^{d \times m}$ are smooth functions, and $(W_t)_{t \geq 0}$ is a $m$-dimensional Brownian motion. This stochastic differential equation is written using the It\={o} convention. For all $x \in \R^d$, the diffusion matrix $\sigma(x)\sigma(x)^{\transp} \in \R^{d \times d}$ is denoted by $a(x)$. Throughout this section, it is assumed to be everywhere nondegenerate, in the sense that for all $x \in \R^d$, for all $\xi \in \R^d\setminus\{0\}$, $\scal{\xi}{a(x)\xi} > 0$.


\subsection{Generalities on the Freidlin-Wentzell theory} We first introduce the Freidlin-Wentzell action functional, and the notion of quasipotential with respect to an equilibrium position of the relaxation dynamics.

\subsubsection{Relaxation dynamics and equilibrium positions} Given $x \in \R^d$, we will call {\em relaxation dynamics} started at $x$ the path $(\psi^x_t)_{t \geq 0}$ solution to the (forward) Cauchy problem
\begin{equation}
  \left\{\begin{aligned}
    & \dot{\psi}^x_t = b(\psi^x_t), \qquad t \geq 0,\\
    & \psi^x_0 = x.
  \end{aligned}\right.
\end{equation}

An {\em equilibrium position} of the relaxation dynamics is a point $\bar{x} \in \R^d$ such that $b(\bar{x})=0$, so that $\psi^{\bar{x}}_t=\bar{x}$ for all $t \geq 0$. A {\em stable equilibrium position} is an equilibrium position $\bar{x} \in \R^d$ such that there exists an open subset $O \subset \R^d$ containing $\bar{x}$ and such that, for all $x \in O$, $\psi^x_t$ converges to $\bar{x}$ when $t$ grows to infinity. Note that this implies that the eigenvalues of the Jacobian matrix $\Diff b(\bar{x})$ have nonpositive real part.

\subsubsection{The Freidlin-Wentzell action functional} When the intensity $\epsilon$ of the noise in~\eqref{eq:SDE} vanishes, the sample-paths of the process $(X_t^{\epsilon})_{t \in [0,T]}$ started at $x \in \R^d$ naturally converge to the relaxation trajectory $(\psi^x_t)_{t \in [0,T]}$, for all finite time $T$. The cornerstone of the Freidlin-Wentzell theory~\cite{FreWen12} is the description of the fluctuations of the diffusion process around this deterministic limit through the {\em action functional}, generically defined for all $T_1 < T_2$ and for all paths $\phi = (\phi_t)_{t \in [T_1,T_2]}$ by
\begin{equation}
  S_{[T_1,T_2],x}(\phi) := \frac{1}{4} \int_{t=T_1}^{T_2} \scal{\dot{\phi}_t - b(\phi_t)}{a(\phi_t)^{-1}(\dot{\phi}_t - b(\phi_t))}\dd t
\end{equation}
if $\phi$ belongs to the set $\ACs([T_1,T_2])$ of absolutely continuous paths such that $\dot{\phi} \in \Ls^2([T_1,T_2])$, and $\phi_{T_1}=x$; and
\begin{equation}
  S_{[T_1,T_2],x}(\phi) := +\infty
\end{equation}
otherwise. Then the Freidlin-Wentzell large deviation principle is the logarithmic equivalence
\begin{equation}
  \forall T>0, \qquad \Pr\left[\forall t \in [0,T], X^{\epsilon}_t \simeq \phi_t\right] \asymp \exp\left(-\frac{S_{[0,T],x}(\phi)}{\epsilon}\right).
\end{equation}
We refer to~\cite{FreWen12, DemZei10} for a mathematical introduction to large deviation theory and for a rigorous formulation of this principle.  

\subsubsection{Quasipotential} The convergence of the diffusion process to the relaxation dynamics is naturally expressed by the fact that $S_{[0,T],x}(\psi^x)$ vanishes for all $T>0$. In fact, the action functional measures the difficulty for the diffusion process to deviate from its typical behaviour. This is best observed by introducing the {\em quasipotential} $V(\bar{x},x)$ with respect to a stable equilibrium position $\bar{x} \in \R^d$, defined for all $x \in \R^d$ by
\begin{equation}\label{eq:qp}
  V(\bar{x},x) := \inf\{S_{[T_1,T_2],\bar{x}}(\phi) : \phi_{T_1} = \bar{x}, \phi_{T_2} = x, T_1<T_2\}.
\end{equation}
Note that $V(\bar{x},\bar{x})=0$, $V(\bar{x},x) \geq 0$ for all $x \in \R^d$, and the function $x \mapsto V(\bar{x},x)$ is continuous on $\R^d$.


\subsection{Transverse decomposition of the drift} In this subsection, we show that the quasipotential actually behaves like a potential for the irreversible dynamics.

\subsubsection{Fluctuation dynamics and identification of the quasipotential}\label{sss:qp} In this paragraph, we recall the framework of~\cite[Section~4.3]{FreWen12}, where the quasipotential can be explicitly identified. Let us fix an open subset $D$ of $\R^d$ such that $\scal{b(y)}{n(y)} < 0$ for all $y \in \partial D$, where $n(y)$ refers to the outward normal vector to $\partial D$. The boundary is said to be {\em noncharacteristic}. This assumption ensures that the relaxation dynamics started at $x \in D \cup \partial D$ cannot exit $D$.

We furthermore assume that there exist smooth functions $U : D \cup \partial D \to \R$ and $\ell : D \cup \partial D \to \R^d$ such that:
\begin{enumerate}[label=(\roman*), ref=\roman*]
  \item for all $x \in D$, $b(x) = -a(x) \nabla U(x) + \ell(x)$ and $\scal{\nabla U(x)}{\ell(x)}=0$,
  \item there exists $\bar{x} \in D$ such that, for all $x \in D\setminus\{\bar{x}\}$, $U(x) > U(\bar{x})$ and $\nabla U(x) \not= 0$.
\end{enumerate}
The function $U$ will be referred to as the {\em potential}. It can already be noted that, for all $x \in D \cup \partial D$,
\begin{equation}
  \frac{\dd}{\dd t} U(\psi^x_t) = - \scal{a(\psi^x_t)\nabla U(\psi^x_t)}{\nabla U(\psi^x_t)} \leq 0,
\end{equation}
so that $U$ is a Lyapunov functional for the relaxation dynamics. As a consequence, when $t$ grows to $+\infty$, $\psi^x_t$ necessarily converges to $\bar{x}$, which is thus a stable equilibrium position of the relaxation dynamics. 

Assuming this {\em transverse decomposition} of the vector field $b$, let us define the {\em fluctuation dynamics} terminated at $x \in D \cup \partial D$ as the path $(\varphi^x_t)_{t \leq 0}$ solving the (backward) Cauchy problem
\begin{equation}
  \left\{\begin{aligned}
    & \dot{\varphi}^x_t = a(\varphi^x_t)\nabla U(\varphi^x_t) + \ell(\varphi^x_t), \qquad t \leq 0,\\
    & \varphi^x_0 = x.
  \end{aligned}\right.
\end{equation}
The fact that this path does not leave $D$ is an assumption that has to be made on the shape of $\partial D$, for example, it is sufficient to assume that $\scal{a(y)\nabla U(y) + \ell(y)}{n(y)} > 0$ for all $y \in \partial D$. When $\ell \equiv 0$, the fluctuation dynamics is the time-reversal of the relaxation dynamics:
\begin{equation}
  \forall t \leq 0, \qquad \varphi^x_t = \psi^x_{-t}.
\end{equation}
In general, some properties related with time-reversal still hold true, for example the potential $U$ remains a (backward) Lyapunov functional for the fluctuation dynamics, in the sense that 
\begin{equation}
  \frac{\dd}{\dd t} U(\varphi^x_t) = \scal{a(\varphi^x_t)\nabla U(\varphi^x_t)}{\nabla U(\varphi^x_t)} \geq 0.
\end{equation}
This implies that, for all $x \in D \cup \partial D$, the fluctuation dynamics terminated at $x$ satisfies 
\begin{equation}
  \lim_{t \to -\infty} \varphi^x_t = \bar{x}.
\end{equation}
It then follows that $\varphi^x$ is an extremal path for~\eqref{eq:qp}, in the sense that
\begin{equation}\label{eq:idqp}
  \begin{aligned}
    V(\bar{x},x) & = U(x) - U(\bar{x})\\
    & = \frac{1}{4} \int_{t=-\infty}^0 \scal{\dot{\varphi}^x_t - b(\varphi^x_t)}{a(\varphi^x_t)^{-1}(\dot{\varphi}^x_t - b(\varphi^x_t))}\dd t,
  \end{aligned}
\end{equation}
see~\cite[Theorem~3.1, p.~100]{FreWen12}. As a consequence, $\varphi^x$ describes the typical trajectory employed by the diffusion process to escape from the neighbourhood of $\bar{x}$ and reach $x$ under the effect of a fluctuation of the noise --- whence the designation {\em fluctuation dynamics}.

\subsubsection{Examples of transverse decompositions} Obviously, any gradient system, or {\em overdamped Langevin process}
\begin{equation}
  \dd X^{\epsilon}_t = - \nabla U(X^{\epsilon}_t)\dd t + \sqrt{2\epsilon} \dd W_t,
\end{equation}
provides a trivial situation where the drift vector field admits a transverse decomposition. In addition to modeling purposes in equilibrium statistical physics, such a stochastic differential equation has a computational interest as simulating its solution over long times allows one to sample from the associated Gibbs measure $\exp(-U/\epsilon)$~\cite{LelRouSto10}. In this perspective, it is known that modifying the drift of the process in order to make it irreversible without changing its stationary measure generically improves its rate of convergence. A means to do so~\cite{HwaHwaShe93, HwaHwaShe05, LelNiePav13} is to consider the stochastic differential equation
\begin{equation}
  \dd X^{\epsilon}_t = - \nabla U(X^{\epsilon}_t)\dd t + J\nabla U(X^{\epsilon}_t)\dd t + \sqrt{2\epsilon} \dd W_t,
\end{equation}
with $J$ being any constant antisymmetric matrix of size $d$. More generally, one can consider the stochastic differential equation
\begin{equation}
  \dd X^{\epsilon}_t = - K\nabla U(X^{\epsilon}_t)\dd t + \sqrt{2\epsilon}\sigma \dd W_t,
\end{equation}
with the symmetric part $\frac{1}{2}(K+K^{\transp})$ of $K$ given by $a=\sigma\sigma^{\transp}$. Then it is easily checked that the Gibbs measure $\exp(-U/\epsilon)$ remains stationary. Besides, letting $\ell(x) := \frac{1}{2}(K-K^{\transp})\nabla U(x)$ yields the relation $\scal{\nabla U(x)}{\ell(x)} = 0$ everywhere. This formulation includes the Kac-Zwanzig model and its generalisation addressed by Ariel and Vanden-Eijnden~\cite{AriVdE07}, as well as kinetic Langevin processes
\begin{equation}
  \left\{\begin{aligned}
    & \dd q_t = p_t \dd t,\\
    & \dd p_t = -\nabla_q u(q_t) \dd t - \gamma p_t \dd t + \sqrt{2\epsilon\gamma} \dd w_t,
  \end{aligned}\right.
\end{equation}
for which $U(q,p) = |p|^2/2 + u(q)$ and the diffusion matrix 
\begin{equation}
  a = \left(\begin{array}{cc}
    0 & 0\\
    0 & \gamma
  \end{array}\right)
\end{equation}
is degenerate.

The examples above have the peculiarity of maintaining the Gibbs measure $\exp(-U/\epsilon)$ stationary for the process, which is due to the fact that $\div \ell \equiv 0$ in these examples, but should not be the case for general irreversible processes. An instance of a vector field with transverse decomposition and for which the stationary distribution does not have an explicit expression is provided by the AB model from~\cite{BouTou12}, which displays similar features to infinite-dimensional models such as 2D Euler equations, the Vlasov equation, magneto-hydrodynamic equations, and the shallow-water equations.

\subsubsection{Hamilton-Jacobi equation and genericity of the transverse decomposition} The relation 
\begin{equation}
  \forall x \in D, \qquad \scal{\nabla U(x)}{\ell(x)}=0
\end{equation}
of the transverse decomposition described above equivalently rewrites
\begin{equation}\label{eq:HJ}
  \forall x \in D, \qquad \scal{\nabla U(x)}{a(x)\nabla U(x)} + \scal{b(x)}{\nabla U(x)} = 0,
\end{equation}
which is the Hamilton-Jacobi equation for the variational problem defining the quasipotential~\eqref{eq:qp}. As a consequence, even in the absence of an explicit transverse decomposition, it can be proved that as soon as the function $x \mapsto V(\bar{x},x)$ is $\Cs^1$, then it is a solution of this Hamilton-Jacobi equation~\cite[Section~4.3]{FreWen12}. Defining $\ell(x) := b(x) + a(x) \nabla_x V(\bar{x},x)$, we deduce that $\scal{\nabla_x V(\bar{x},x)}{\ell(x)}=0$ for all $x \in \R^d$. In other words, under some smoothness assumption on the quasipotential, the latter {\em automatically} provides an implicit transverse decomposition of the vector field $b$. Of course, this transverse decomposition is only useful in domains $D$ satisfying the conditions of~\S\ref{sss:qp}.

Singularities of the quasipotential, that is to say situations in which $x \mapsto V(\bar{x},x)$ fails to satisfy the Hamilton-Jacobi equation~\eqref{eq:HJ}, have been an important topic of research in nonequilibrium physics for several decades~\cite{GraHak71, GraTel84JSP, GraTel85, MaiSte93, DykMilSme94, DykSmeMaiSil96, MaiSte97}, and have known a recent renewed interest due to their link with Lagrangian phase transitions in the macroscopic fluctuation theory of driven diffusive systems. We refer in particular to the recent review~\cite{BaeKaf15} by Baek and Kafri for a detailed survey. In this paper we shall not address this issue and generically assume that the quasipotential is $\Cs^1$ in domains of interest.


\section{Prefactor for the stationary distribution}\label{s:stat}

This section is dedicated to the study of the stationary distribution of the diffusion process $(X^{\epsilon}_t)_{t \geq 0}$ solution to~\eqref{eq:SDE}. We shall therefore work under the standing assumption that for $\epsilon>0$ small enough, this process possesses a unique stationary distribution $\Pst^{\epsilon}$, and that this probability distribution possesses a smooth density $\pst^{\epsilon}$ with respect to the Lebesgue measure on $\R^d$. Practical conditions on the functions $b$ and $\sigma$ ensuring this assumption can be found in~\cite{Kha11} or~\cite{FreWen12}. Then the density $\pst^{\epsilon}$ solves the stationary Fokker-Planck equation
\begin{equation}\label{eq:sFP}
  \forall x \in \R^d, \qquad 0 = \epsilon \sum_{i,j=1}^d \partial_{ij}\left(a_{ij}(x) \pst^{\epsilon}(x)\right) - \sum_{i=1}^d \partial_i\left(b_i(x)\pst^{\epsilon}(x)\right).
\end{equation}


\subsection{WKB approach} When, for all $x \in \R^d$, $a(x)$ is the identity matrix and $b(x) = -\nabla U(x)$ with 
\begin{equation}
  Z^{\epsilon} := \int_{x \in \R^d} \exp\left(-\frac{U(x)}{\epsilon}\right)\dd x < +\infty,
\end{equation}
it is well known that $\Pst^{\epsilon}$ is the {\em Gibbs measure} defined by the density
\begin{equation}\label{eq:Gibbs}
  \pst^{\epsilon}(x) = \frac{1}{Z^{\epsilon}}\exp\left(-\frac{U(x)}{\epsilon}\right).
\end{equation}
Assuming that $U$ attains its global minimum at a unique point $\bar{x} \in \R^d$ and that $\Hess U(\bar{x})$ is positive-definite, the Laplace approximation of $Z^{\epsilon}$ yields the equivalence
\begin{equation}\label{eq:pstpot}
  \pst^{\epsilon}(x) \Sim_{\epsilon \dto 0} \sqrt{\frac{\det \Hess U(\bar{x})}{(2\pi\epsilon)^d}}\exp\left(-\frac{U(x)-U(\bar{x})}{\epsilon}\right).
\end{equation}

In the general case, the WKB analysis of the stationary Fokker-Planck equation~\eqref{eq:sFP} consists in looking for a solution of the form
\begin{equation}
  \pst^{\epsilon}(x) = \frac{\Cst^{\epsilon}(x)}{\epsilon^{d/2}} \exp\left(-\frac{U(x)}{\epsilon}\right),
\end{equation}
where the prefactor $\Cst^{\epsilon}(x)/\epsilon^{d/2}$ is subexponential. This approach has been widely used~\cite{Gra88,MatSch77,MaiSte97,Sch09} and it is known that injecting the ansatz above in~\eqref{eq:sFP} and identifying the terms according to the powers of $\epsilon$ yields the Hamilton-Jacobi equation~\eqref{eq:HJ} for $U$, and the transport equation
\begin{equation}\label{eq:transport}
  \scal{\nabla \Cst}{b + 2a\nabla U} + \Cst \left(\div b + a:\Hess U + 2\scal{A}{\nabla U}\right) = 0,
\end{equation}
for the lowest order approximation of the prefactor
\begin{equation}
  \Cst(x) := \lim_{\epsilon \dto 0} \Cst^{\epsilon}(x).
\end{equation}
In the transport equation~\eqref{eq:transport}, the vector $A(x) \in \R^d$ is defined by
\begin{equation}
  A_i(x) := \sum_{j=1}^d \partial_j a_{ij}(x).
\end{equation} 


\subsection{Quasipotential in the eikonal equation} In the context of WKB approximation, the Hamilton-Jacobi equation~\eqref{eq:HJ} for $U$ is generally referred to as the {\em eikonal} equation. If the relaxation dynamics possesses a unique stable equilibrium position $\bar{x} \in \R^d$ such that $x \mapsto V(\bar{x},x)$ is $\Cs^1$ on $\R^d$ and $\nabla_x V(\bar{x},x) \not= 0$ for $x \not= \bar{x}$, then the discussion of~\S\ref{sss:qp} shows that $V(\bar{x},x)=U(x)-U(\bar{x})$ for all $x \in \R$. A probabilistic formulation of this result is the fact that the family of probability distributions $\{\Pst^{\epsilon}; \epsilon > 0\}$ satisfies a large deviation principle with rate function $V(\bar{x},x)$~\cite[Theorem~4.3 in Section~4.4]{FreWen12}. 


\subsection{Explicit solution of the transport equation} Under the assumptions of~\S\ref{sss:qp}, let us denote $\ell(x) := b(x) + a(x)\nabla U(x)$, with $\nabla U(x) = \nabla_x V(\bar{x},x)$. Then the transport equation~\eqref{eq:transport} rewrites
\begin{equation}
  \scal{\nabla \Cst}{a\nabla U + \ell} + \Cst \left(\div \ell + \scal{A}{\nabla U}\right) = 0.
\end{equation}

In the WKB approach, this equation is usually solved by the method of characteristics, or {\em rays}~\cite{CohLew67,Lud75}. The main interest of the formulation of the problem in terms of large deviation theory is that these rays are exactly given by the trajectories of the fluctuation dynamics. Indeed, the equation above rewrites
\begin{equation}
  \scal{\nabla \log\Cst}{a\nabla U + \ell} = -F,
\end{equation}
where 
\begin{equation}
  F(x) := \div \ell(x) + \scal{A(x)}{\nabla U(x)}.
\end{equation}
We note that if $F \equiv 0$, then the Gibbs measure defined by~\eqref{eq:Gibbs} is stationary for the process, although the latter need not be reversible. In other words, $F$ is a measure of {\em non-Gibbsianness} of the system.

Recalling that $a\nabla U + \ell$ is the vector field driving the fluctuation dynamics, we obtain that, for all $x \in \R^d$, for all $t \leq 0$,
\begin{equation}
  \frac{\dd}{\dd t} \log \Cst(\varphi^x_t) = \scal{\nabla \log\Cst(\varphi^x_t)}{a(\varphi^x_t)\nabla U(\varphi^x_t) + \ell(\varphi^x_t)} = -F(\varphi^x_t),
\end{equation}
so that integrating this identity for $t \in (-\infty,0]$ finally yields
\begin{equation}\label{eq:Cst:1}
  \Cst(x) = \Cst(\bar{x})\exp\left(- \int_{t=-\infty}^0 F(\varphi^x_t)\dd t\right).
\end{equation}
Under the assumption that $\Hess U(\bar{x}) = \Hess_x V(\bar{x},\bar{x})$ be positive-definite, performing a Laplace approximation in the normalisation condition for $\pst^{\epsilon}$ yields
\begin{equation}\label{eq:Cst:2}
  \Cst(\bar{x}) = \sqrt{\frac{\det \Hess U(\bar{x})}{(2\pi)^d}},
\end{equation}
and we finally come up with the formula
\begin{equation}\label{eq:st}
  \pst^{\epsilon}(x) \Sim_{\epsilon \dto 0} \sqrt{\frac{\det \Hess_x V(\bar{x},\bar{x})}{(2\pi\epsilon)^d}} \exp\left(-\frac{V(\bar{x},x)}{\epsilon} - \int_{t=-\infty}^0 F(\varphi^x_t)\dd t\right)
\end{equation}
for the stationary density. It is striking that it only differs from the formula~\eqref{eq:pstpot} for the potential case through the integral term along the fluctuation dynamics, which therefore has to be interpreted as the accumulation along the fluctuation dynamics of some non-Gibbsianness of the system, measured by the function $F$.


\section{Quasistationary exit rate of a domain}\label{s:bl}

In this section, we consider an open subset $D \subset \R^d$ with noncharacteristic boundary $\partial D$ and such that all the trajectories of the relaxation dynamics started in $D \cup \partial D$ converge to a unique stable equilibrium position $\bar{x} \in D$. As a consequence, when $\epsilon$ is small, the diffusion process $X^{\epsilon}$ typically fluctuates around the point $\bar{x}$, and this process hitting $\partial D$ is a rare event. To quantify its time scale, let us introduce
\begin{equation}
  \tau^{\epsilon}_{\partial D} := \inf\{t>0, X^{\epsilon}_t \in \partial D\}.
\end{equation}
The Freidlin-Wentzell theory~\cite[Section~4.4]{FreWen12} asserts that $\tau^{\epsilon}_{\partial D}$ is of order $\exp(V(\bar{x},\partial D)/\epsilon)$, where $V(\bar{x},\partial D) := \min_{y \in \partial D} V(\bar{x},y)$. We therefore define the {\em quasistationary phase} as the range of times
\begin{equation}\label{eq:locth}
  1 \ll t \ll \exp\left(\frac{V(\bar{x},\partial D)}{\epsilon}\right),
\end{equation}
during which the process forgets its initial position but typically remains stuck in a neighbourhood of $\bar{x}$.

The purpose of this section is to obtain a formula for the exit rate from $D$ during the quasistationary phase. This exit rate is defined in Subsection~\ref{ss:exit}. It relies on the solution to the stationary Fokker-Planck equation in $D$ with absorbing boundary condition, which is well approximated in the range of times~\eqref{eq:locth} by the {\em quasistationary distribution} introduced in Subsection~\ref{ss:qsd}. The exit rate is computed through a boundary layer approximation for the quasistationary distribution in Subsection~\ref{ss:bl}.


\subsection{Probability current and exit rate}\label{ss:exit} Assume that the particle is absorbed, or killed, when it reaches the boundary of $D$. Then the law of its position, defined by
\begin{equation}
  \forall t \geq 0, \quad \forall x \in D, \qquad \pabs^{\epsilon}(t,x) := \Pr\left[X^{\epsilon}_t \simeq x, \tau^{\epsilon}_{\partial D} > t\right],
\end{equation}
satisfies the Fokker-Planck equation
\begin{equation}\label{eq:FPabs}
  \partial_t \pabs^{\epsilon} = \epsilon \sum_{i,j=1}^d \partial_{i,j}\left(a_{ij}\pabs^{\epsilon}\right) - \sum_{i=1}^d \partial_i\left(b_i\pabs^{\epsilon}\right), \qquad t \geq 0, \quad x \in D,
\end{equation}
supplemented with the absorbing boundary condition 
\begin{equation}\label{eq:babs}
  \pabs^{\epsilon}(t,x) = 0, \qquad x \in \partial D,
\end{equation}
see~\cite[Section~5.3.2]{Gar85}. 

There is of course a loss of probability as time evolves, in the sense that 
\begin{equation}
  \int_{x \in D} \pabs^{\epsilon}(t,x) \dd x = \Pr\left[\tau^{\epsilon}_{\partial D} > t\right]
\end{equation}
decreases, corresponding to the probability that the particle exits $D$. This loss is measured by the {\em probability current} $j^{\epsilon}(t,x)$, defined by the formulation of the Fokker-Planck equation~\eqref{eq:FPabs} as the conservation law
\begin{equation}
  \partial_t \pabs^{\epsilon}(t,x) + \div j^{\epsilon}(t,x) = 0,
\end{equation}
so that
\begin{equation}\label{eq:j}
  j^{\epsilon}(t,x) = -\epsilon\left(A(x)\pabs^{\epsilon} + a(x) \nabla \pabs^{\epsilon}\right) + b(x) \pabs^{\epsilon},
\end{equation}
for all $t \geq 0$ and $x \in D$. Then, the rate of exit from $D$ is given by the flux of $j^{\epsilon}$ through $\partial D$, namely
\begin{equation}
  \lambda^{\epsilon}(t) := \int_{y \in \partial D} \scal{j^{\epsilon}(t,y)}{n(y)} \dd S(y),
\end{equation}
where $\dd S(y)$ is the surface element and $n(y)$ is the outward normal to $\partial D$, see for instance~\cite[Sections~2.5 and~2.6]{MaiSte97},~\cite[Section~10.2]{Sch09} or~\cite[Section~5.4]{Gar85}.


\subsection{Quasistationary distribution}\label{ss:qsd} Let us define the {\em quasistationary distribution} $\pqst^{\epsilon}$ associated with $D$ by the so-called Yaglom limit of the conditional probability
\begin{equation}
  \pqst^{\epsilon}(x) := \lim_{t \to +\infty} \Pr\left[X^{\epsilon}_t \simeq x | \tau^{\epsilon}_{\partial D} > t\right] = \lim_{t \to +\infty} \frac{\pabs^{\epsilon}(t,x)}{\Pr\left[\tau^{\epsilon}_{\partial D} > t\right]},
\end{equation}
so that for $t \gg 1$, $\pabs^{\epsilon}(t,x) \simeq \pqst^{\epsilon}(x)\Pr\left[\tau^{\epsilon}_{\partial D} > t\right]$. On the other hand, on the time scale~\eqref{eq:locth}, $\Pr\left[\tau^{\epsilon}_{\partial D} > t\right]$ remains close to $1$. As a consequence, during the quasistationary phase, the distribution $\pabs^{\epsilon}(t,x)$ is well approximated by the quasistationary distribution $\pqst^{\epsilon}(x)$, and in particular it does not vary in time.

During this phase, the exit rate thus writes
\begin{equation}\label{eq:lqst}
  \lqst^{\epsilon} := \int_{y \in \partial D} \scal{\jqst^{\epsilon}(y)}{n(y)} \dd S(y),
\end{equation}
where the {\em quasistationary current} $\jqst^{\epsilon}$ is defined by approximating $\pabs^{\epsilon}(t,x)$ with $\pqst^{\epsilon}(x)$ in~\eqref{eq:j}. On the boundary of $D$, the absorbing condition reduces its expression to
\begin{equation}\label{eq:jdD}
  \forall y \in \partial D, \qquad \jqst^{\epsilon}(y) = - \epsilon a(y) \nabla \pqst^{\epsilon}(y).
\end{equation}
We compute this quasistationary current in the next subsection.


\subsection{Boundary layer approximation and computation of the quasistationary current}\label{ss:bl} The purpose of this paragraph is to compute the quasistationary current $\jqst^{\epsilon}$ on $\partial D$, in order to derive a formula for $\lqst^{\epsilon}$. 

\subsubsection{Assumptions on the domain} We assume that the quasipotential $x \mapsto V(\bar{x},x)$ is $\Cs^1$, and that, in addition to the fact that the boundary $\partial D$ be noncharacteristic, it satisfies the condition that for all $y \in \partial D$,
\begin{equation}\label{eq:asspartialD}
  \scal{-a(y)\nabla_x V(\bar{x},y) + \ell(y)}{n(y)} < 0 \quad \text{and} \quad \scal{\nabla_x V(\bar{x},y)}{n(y)} > 0,
\end{equation}
where we have defined $\ell(x) := b(x) + a(x)\nabla_x V(\bar{x},x)$ and $n(y)$ is the outward normal vector at $y \in \partial D$. This ensures that the results of~\S\ref{sss:qp} hold with $U(x) = V(\bar{x},x)$, up to an additive constant.

\subsubsection{Boundary layer approximation} Conditioning the sample-paths of $X^{\epsilon}$ with respect to the event $\{\tau^{\epsilon}_{\partial D} > t\}$ in the definition of the quasistationary distribution induces a repulsive effect of the boundary on the particle. When the latter is far from $\partial D$, it does not feel this effect and therefore the quasistationary distribution in the bulk of $D$ can be approximated by the expression~\eqref{eq:st} of the stationary distribution of a particle diffusing in a single-well potential given by the restriction of $U$ to $D$. In the sequel, this distribution is referred to as the {\em ensemble measure} 
\begin{equation}
  p^{\epsilon}_{\mathrm{ens}} = \frac{\Cst(x)}{\epsilon^{d/2}} \exp\left(-\frac{V(\bar{x},x)}{\epsilon}\right),
\end{equation}
where we recall that $\Cst(x)$ is defined by~\eqref{eq:Cst:1} and~\eqref{eq:Cst:2}.

On the other hand, the repulsion effect makes the quasistationary density vanish in the neighbourhood of $\partial D$, where we therefore have to perform a boundary layer approximation to obtain the expression of $\pqst^{\epsilon}$~\cite{MatSch77,MaiSte97,Sch09}. The thickness of this layer is determined by the fact that boundary layer effects must be caused by a subexponential prefactor; in other words, if $x \in D$ is close to $\partial D$ and $y \in \partial D$ denotes its orthogonal projection on $\partial D$, then $x$ is out of the boundary layer as soon as $\exp(-V(\bar{x},x)/\epsilon) \gg \exp(-V(\bar{x},y)/\epsilon)$. Writing, for $|x-y| \ll 1$,
\begin{equation}
  \begin{aligned}
    V(\bar{x},x) & \simeq V(\bar{x},y) + \scal{\nabla_x V(\bar{x},y)}{x-y}\\
         & \simeq V(\bar{x},y) - |x-y|\scal{\nabla_x V(\bar{x},y)}{n(y)},
  \end{aligned}
\end{equation}
and using the assumption~\eqref{eq:asspartialD} that $\scal{\nabla_x V(\bar{x},y)}{n(y)} > 0$, we deduce that $\exp(-V(\bar{x},x)/\epsilon) \gg \exp(-V(\bar{x},y)/\epsilon)$ as soon as $|x-y| \gg \epsilon$, so that the boundary layer develops on a characteristic length scale $\epsilon$.

This discussion leads us to assume that in the vicinity of $\partial D$, the quasistationary distribution writes
\begin{equation}\label{eq:quasistat}
  \pqst^{\epsilon}(x) = \frac{\Cbl(\pi_D(x),\eta(x)/\epsilon)}{\epsilon^{d/2}} \exp\left(-\frac{V(\bar{x},x)}{\epsilon}\right),
\end{equation}
where $\pi_D(x)$ is the orthogonal projection of $x$ onto $\partial D$, $\eta(x)$ is the distance between $x$ and $\pi_D(x)$. Matching the quasistationary distribution with the ensemble measure $p^{\epsilon}_{\mathrm{ens}}$ in the bulk of $D$ yields the boundary condition
\begin{equation}\label{eq:bcCbl}
  \forall y \in \partial D, \qquad \lim_{r \to +\infty} \Cbl(y,r) = \Cst(y).
\end{equation}

We point out the fact that the ansatz~\eqref{eq:quasistat} is well suited to determine the leading order approximation of the quasistationary distribution, which is enough for our purpose. If one wants to obtain corrections at next orders, one should rather assume the prefactor to be a function of both $x$ and $\eta(x)$. This would lead to compatibility conditions to be solved.
 
\subsubsection{Computation of $\jqst^{\epsilon}(y)$} Differentiating the expression~\eqref{eq:quasistat} with respect to $x$ and injecting the result into the expression~\eqref{eq:jdD} of the quasistationary current on $\partial D$, we obtain
\begin{equation}
  \jqst^{\epsilon}(y) \simeq - \frac{1}{\epsilon^{d/2}}a(y) \nabla \eta(y) \frac{\partial \Cbl}{\partial r}(y,0) \exp\left(-\frac{V(\bar{x},y)}{\epsilon}\right)
\end{equation}
at the lowest order in $\epsilon$. Identifying $\nabla \eta(y)$ with $-n(y)$, we conclude that
\begin{equation}
  \jqst^{\epsilon}(y) = \frac{1}{\epsilon^{d/2}} a(y) n(y) \frac{\partial \Cbl}{\partial r}(y,0) \exp\left(-\frac{V(\bar{x},y)}{\epsilon}\right).
\end{equation}

We now compute $\frac{\partial \Cbl}{\partial r}(y,0)$. In this purpose, we fix $y \in \partial D$, $r>0$ and let $x := y - \epsilon r n(y) \in D$. Then injecting the expression~\eqref{eq:quasistat} of $\pqst^{\epsilon}(x)$ into the Fokker-Planck equation~\eqref{eq:FPabs} yields, at the lowest order in $\epsilon$,
\begin{equation}\label{eq:odeCbl}
  \scal{n(y)}{a(y)n(y)} \frac{\partial^2 \Cbl}{\partial r^2}(y,r) + \scal{a(y)\nabla_x V(\bar{x},y) + \ell(y)}{n(y)} \frac{\partial \Cbl}{\partial r}(y,r) = 0.
\end{equation}
Note that the nondegeneracy assumption on $a$ implies that $\scal{n(y)}{a(y)n(y)}>0$. Let 
\begin{equation}
  \mu(y) := \frac{\scal{a(y)\nabla_x V(\bar{x},y) + \ell(y)}{n(y)}}{\scal{n(y)}{a(y)n(y)}}.
\end{equation}
By~~\eqref{eq:asspartialD}, $\mu(y)>0$. As a consequence, the equation~\eqref{eq:odeCbl} integrates as
\begin{equation}
  \Cbl(y,r) = \Cbl(y,0) + \frac{\partial \Cbl}{\partial r}(y,0)\frac{1-\expo{-\mu(y)r}}{\mu(y)},
\end{equation}
and the boundary conditions $\Cbl(y,0)=0$, which follows from~\eqref{eq:babs}, as well as~\eqref{eq:bcCbl} imply
\begin{equation}
  \frac{\partial \Cbl}{\partial r}(y,0) = \mu(y) \Cst(y).
\end{equation}
We deduce that the quasistationary current writes
\begin{equation}\label{eq:jsqt}
  \jqst^{\epsilon}(y) = \frac{\Cst(y)}{\epsilon^{d/2}} \exp\left(-\frac{V(\bar{x},x)}{\epsilon}\right) \frac{\scal{a(y)\nabla_x V(\bar{x},y) + \ell(y)}{n(y)}}{\scal{n(y)}{a(y)n(y)}} a(y)n(y).
\end{equation}

\subsubsection{Conclusion} Injecting the expression of the quasistationary current into the definition~\eqref{eq:lqst} of the quasistationary exit rate yields
\begin{equation}
  \lqst^{\epsilon} = \int_{y \in \partial D} \frac{\Cst(y)}{\epsilon^{d/2}} \exp\left(-\frac{V(\bar{x},y)}{\epsilon}\right) \scal{a(y)\nabla_x V(\bar{x},y) + \ell(y)}{n(y)} \dd S(y).
\end{equation}
One recognises here the expression of the ensemble measure $p^{\epsilon}_{\mathrm{ens}}$ on the one hand, and of the vector field $a\nabla_x V + \ell$ driving the fluctuation dynamics on the other hand, so that the quasistationary exit rate rewrites
\begin{equation}
  \lqst^{\epsilon} = \int_{y \in \partial D} \Scal{{\dot{\varphi}^y_t}|_{t=0}}{n(y)} p_{\mathrm{ens}}^{\epsilon}(y)\dd S(y).
\end{equation}
This brings forth the final interpretation that the escape of the particle from $D$ is governed by the flux through $\partial D$ of the fluctuation dynamics, weighted by the ensemble measure $p^{\epsilon}_{\mathrm{ens}}$ of the particle diffusing in $D$. 


\section{The irreversible Eyring-Kramers formula}\label{s:EK}

We finally address the situation where the process $X^{\epsilon}$ is metastable in the neighbourhood of isolated attractors, and compute the average transition time between these attractors. As is discussed in the introduction of this article, our analysis is local and can therefore be reduced to the simple situation of a bistable system. We therefore assume that the relaxation dynamics possesses two equilibrium points $\bar{x}_1$ and $\bar{x}_2$, the respective basins of attraction of which are separated by a smooth hypersurface $S$ containing a unique saddle-point $x_{\star}$. As is underlined in~\cite{MaiSte97}, the separating surface $S$ is {\em characteristic} in the sense that $\scal{b(y)}{n(y)} = 0$ for all $y \in S$, which makes the boundary layer approach developed in the previous section more delicate, see also~\cite{AriVdE07,Sch09}. We will therefore not follow this approach directly, and rather proceed as follows. 

Consider a particle with initial position $\bar{x}_1$ and evolving according to the diffusion process~\eqref{eq:SDE}. We first use the Freidlin-Wentzell theory to show that, when $\epsilon$ is small, the point at which the particle hits $S$ is concentrated around $x_{\star}$. This allows us to look at the linearised version of the dynamics in the neighbourhood of the saddle-point. For this dynamics, the separating surface between the two attractors is the tangent hyperplane to $S$ at $x_{\star}$, which we denote by $S_0$. For $\eta \ll 1$, we denote by $S_{\eta}$ the hyperplane parallel to $S_0$ and located at a distance of order $\eta$ of the latter, in the direction of $\bar{x}_1$, so that a particle starting from $\bar{x}_1$ needs to cross $S_{\eta}$ before reaching $S_0$. We then compute independently:
\begin{enumerate}[label=(\roman*), ref=\roman*]
  \item the rate at which particles arrive on $S_{\eta}$,
  \item the probability that a particle started from $S_{\eta}$ drifts away from the saddle-point toward $\bar{x}_2$ instead of $\bar{x}_1$.
\end{enumerate}
Matching these two quantities yields the rate at which a particle started from $\bar{x}_1$ goes to the neighbourhood of $\bar{x}_2$, which is the inverse of the average transition time we are willing to compute.


\subsection{Bistability and instanton}\label{ss:bistab} We first describe the bistability of the system and introduce a few notations.

\subsubsection{Vector field $b$} 
\begin{figure}[ht]
    \begin{pspicture}(6,3)
      \rput(3,1.5){\includegraphics[clip=true,trim=3cm 4cm 2.8cm 4cm,width=6cm]{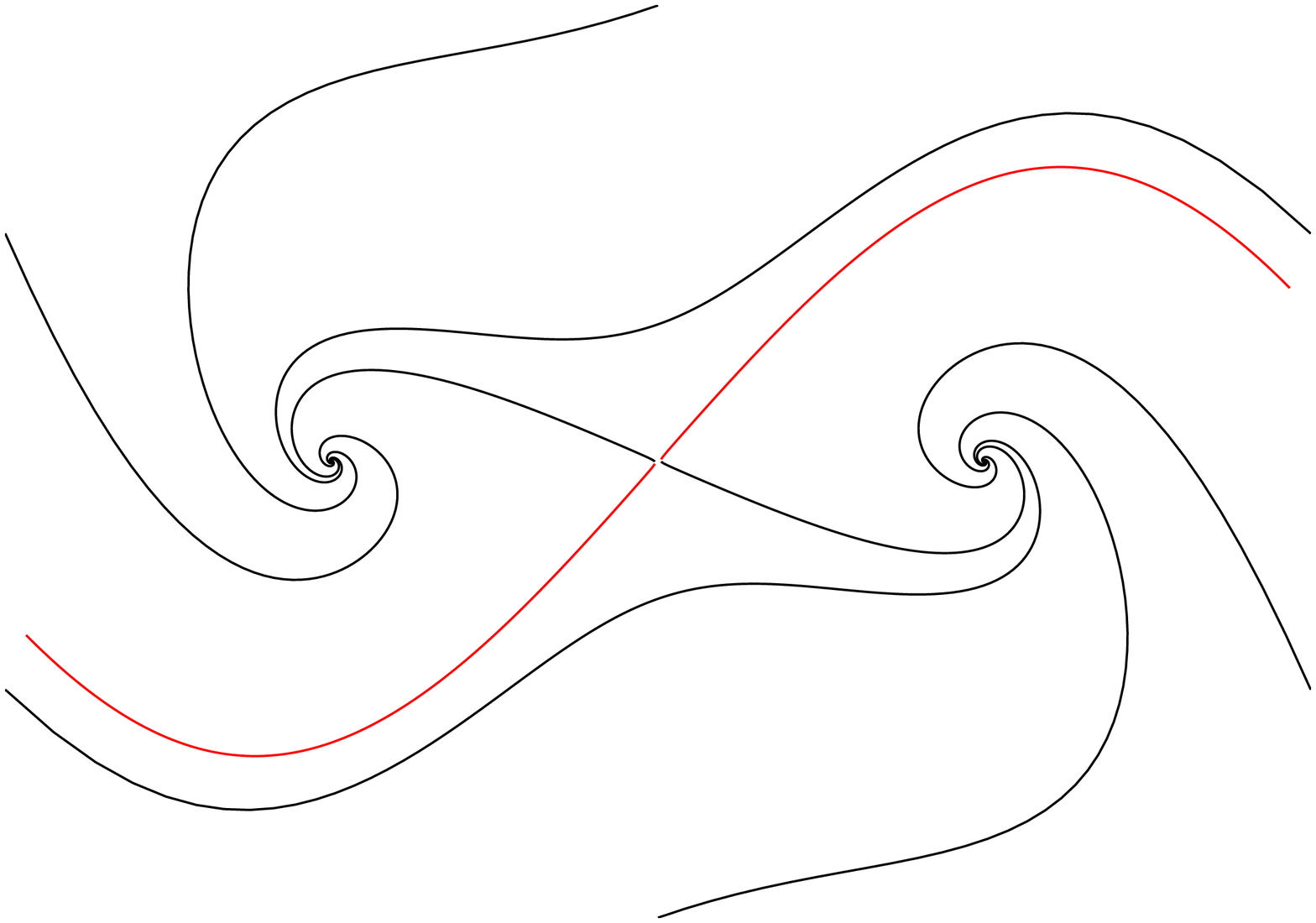}}
      
      \rput(2.97,1.5){\red$\bullet$}
      \rput(3.05,1.3){\red$x_{\star}$}
      \psline[linecolor=red]{->}(2,.53)(2.05,.575)
      \psline[linecolor=red]{->}(3.95,2.47)(3.9,2.425)
      \rput(6,2){\red$S$}
      
      \psline{->}(2.35,.53)(2.4,.575)
      \psline{->}(3.6,2.47)(3.55,2.425)

      \psline{->}(5,.7)(5,.75)
      \psline{->}(.95,2.3)(.95,2.25)
      
      \psline{->}(5.5,1.1)(5.46,1.17)
      \psline{->}(.44,1.9)(.48,1.83)
      
      \psline{->}(2.5,1.7)(2.45,1.72)
      \psline{->}(3.45,1.3)(3.5,1.28)
      
      \rput(1.55,1.5){$\bullet$}
      \rput(1.7,1.7){$\bar{x}_1$}

      \rput(4.4,1.5){$\bullet$}
      \rput(4.25,1.3){$\bar{x}_2$}

    \end{pspicture}
    \caption{Flow lines of the relaxation dynamics for a bistable system.}
    \label{fig:bistab}
\end{figure}

The relaxation dynamics is assumed to possess exactly two stable equilibrium positions $\bar{x}_1$ and $\bar{x}_2$, with respective basins of attraction $D_1$ and $D_2$ separated by a hypersurface $S = \partial D_1 = \partial D_2$. This hypersurface is stable for the relaxation dynamics, and we assume that, for all $x \in S$, $\psi^x_t$ converges to the equilibrium position $x_{\star} \in S$, which is therefore stable in $d-1$ directions and unstable in the remaining direction. See Figure~\ref{fig:bistab}.

We first prove that
\begin{equation}
  V(\bar{x}_1,x_{\star}) = \inf_{y \in S} V(\bar{x}_1,y),
\end{equation}
so that $x_{\star}$ behaves like a saddle-point for the quasipotential. To this aim, we fix $y \in S$ and construct a continuous path $(\phi_t)_{t \in \R}$ as follows:
\begin{itemize}
  \item the part $(\phi_t)_{t \leq 0}$ is an arbitrary continuous path joining $\bar{x}_1$ to $y$,
  \item the part $(\phi_t)_{t \geq 0}$ is given by the relaxation dynamics of from $y$ to $x_{\star}$.
\end{itemize}
Then the part corresponding to the relaxation dynamics does not contribute to the action of $(\phi_t)_{t \in \R}$, so that taking the infimum over all such paths we deduce that $V(\bar{x}_1,y) \geq V(\bar{x}_1, x_{\star})$.

As a consequence, when $\epsilon$ is small, the diffusion process typically fluctuates in the neighbourhood of a local equilibrium for long times, and then jumps to the neighbourhood of the other local equilibrium. If the quasipotential reaches its minimum on $S$ only at the saddle-point, then the typical trajectory employed for this transition crosses $S$ in the neighbourhood of $x_{\star}$.

\subsubsection{Geometry of the dynamics around the saddle-point} Let us denote $\Ms := \Diff b(x_{\star})$ the Jacobian matrix of $b$ at $x_{\star}$. On account of the assumptions on the vector field $b$ around $x_{\star}$, the matrix $\Ms$ has $d-1$ eigenvalues with negative real part, and one nonnegative real eigenvalue. The corresponding stable spaces satisfy the following properties.
\begin{enumerate}[label=(\roman*), ref=\roman*]
  \item There exists a hyperplane $V_-$ such that $\Ms V_- \subset V_-$ and, for all $x \in V_-$, $\exp(t\Ms)x$ converges to $0$ when $t \to +\infty$. The tangent hyperplane $S_0$ to $S$ at $x_{\star}$ writes $S_0 = x_{\star} + V_-$.
  \item The nonnegative real eigenvalue is denoted by $\lambda^{\star}_+$, it is assumed to be positive. The corresponding unit eigenvector is denoted by $v_+$, we prescribe it to be oriented toward $D_2$.
\end{enumerate}
The normal vector to $V_-$ pointing in the direction of $D_2$ is denoted $\ns$, its angle with $v_+$ is denoted $\theta$.  Given $x \in \R^d$, let us define $\zeta_+(x)$ as the unique real number such that $x-x_{\star}-\zeta_+(x)v_+ \in V_-$. We note that, for all $x \in \R^d$,
\begin{equation}\label{eq:zetap}
  \zeta_+(x) = \frac{\scal{x-x_{\star}}{\ns}}{\cos\theta}.
\end{equation}
These notations are summarised on Figure~\ref{fig:saddle:0}.

\subsubsection{Saddle-point and instanton} The {\em instanton} is the typical trajectory employed by the diffusion process to reach the saddle-point starting from $\bar{x}_1$. In general, it takes an infinite amount of time to leave $\bar{x}_1$, and an infinite amount of time to reach $x_{\star}$, so that the instanton has to be though of as a trajectory $(\rho_t)_{t \in \R}$ such that
\begin{equation}
  \lim_{t \to -\infty} \rho_t = \bar{x}_1, \qquad \lim_{t \to +\infty} \rho_t = x_{\star},
\end{equation}
for which the action
\begin{equation}
  \frac{1}{4} \int_{t=-\infty}^{+\infty} \scal{\dot{\rho}_t - b(\rho_t)}{a(\rho_t)^{-1}(\dot{\rho}_t - b(\rho_t))}\dd t
\end{equation}
is minimal and worth $V(\bar{x}_1,x_{\star})$. In the sequel, we shall make the following assumptions:
\begin{enumerate}[label=(\roman*), ref=\roman*]
  \item there is a unique such path (up to time translations);
  \item for all $t \in \R$, the quasipotential $V(\bar{x}_1,\cdot)$ is smooth in the neighbourhood of $\rho_t$, so that it satisfies the Hamilton-Jacobi relation 
  \begin{equation}
    \scal{\nabla_x V(\bar{x}_1,x)}{\nabla_x V(\bar{x}_1,x)} + \scal{b(x)}{\nabla_x V(\bar{x}_1,x)} = 0;
  \end{equation}
  \item when $t \to +\infty$, $\Hess_x V(\bar{x}_1,\rho_t)$ converges to some symmetric nondegenerate matrix $\Hs$ with $d-1$ positive eigenvalues and one negative eigenvalue.
\end{enumerate}
Defining $U(x) := V(\bar{x}_1,x)$ and $\ell(x) := b(x) + a(x)\nabla U(x)$ in the neighbourhood of each $\rho_t$, we deduce from the same arguments as in Section~\ref{s:qp} that $(\rho_t)_{t \geq 0}$ is a heterocline trajectory of the dynamical system
\begin{equation}
  \dot{\rho}_t = a(\rho_t) \nabla U(\rho_t) + \ell(\rho_t),
\end{equation}
connecting the equilibrium positions $\bar{x}_1$ and $x_{\star}$. In particular, for all $t \in \R$, the fluctuation trajectory terminated at $\rho_t$ coincides with the instanton in the sense that
\begin{equation}
  \forall s \leq 0, \qquad \varphi^{\rho_t}_s = \rho_{s+t}.
\end{equation}

Differentiating the transverse relation $\scal{\nabla U(x)}{\ell(x)} = 0$ twice and evaluating the result in $\rho_t$ for $t \to +\infty$ yields
\begin{equation}\label{eq:HDDtH}
  \Hs\Ds + \Ds^{\transp}\Hs = 0,
\end{equation}
where $\Ds := \lim_{t \to +\infty} \Diff \ell(\rho_t)$ satisfies
\begin{equation}
  \Ms = -\as\Hs+\Ds, \qquad \as:=a(x_{\star}).
\end{equation}
We finally note for further purpose that multiplying~\eqref{eq:HDDtH} by $\Hs^{-1}$ on both sides yields
\begin{equation}\label{eq:DHHDt}
  \Ds\Hs^{-1} + \Hs^{-1}\Ds^{\transp} = 0.
\end{equation}

\subsubsection{Linearisation of the fluctuation dynamics} 
\begin{figure}[ht]
  \centering
  \psset{unit=.9cm}
  \begin{pspicture}(12,10)
    \pscustom[linecolor=lightgray,hatchcolor=lightgray,fillstyle=vlines]{\psline(5,10)(6,5)(7,10)}
    
    \psline[linecolor=red,linewidth=2pt]{-}(5,0)(7,10)
    \psline[linecolor=red,ArrowInside=->](6,5)(0,8)
    \psline[linecolor=red,ArrowInside=->](6,5)(12,2)
    \psline[linecolor=red,linewidth=2pt]{->}(6,5)(8.3,3.85)
    \psline[linecolor=red,linestyle=dashed]{->}(6,5)(8.5,4.5)
    \psline[linecolor=red,linestyle=dashed]{-}(6.1,5.5)(6.6,5.4)(6.5,4.9)
    \psarc[linecolor=red,linestyle=dashed](6,5){2}{-26.56}{-11.3}
    
    \pscurve[linecolor=red](4.5,0)(4,4)(0,7)
    \psline[linecolor=red]{->}(4,4)(3.99,4.02)
    \pscurve[linecolor=red](7.5,10)(8,6)(12,3)
    \psline[linecolor=red]{->}(8,6)(8.01,5.98)
    
    \pscurve[linecolor=red](6.5,10)(5,8)(0,9)
    \psline[linecolor=red]{->}(5,8)(4.98,7.99)
    \pscurve[linecolor=red](5.5,0)(7,2)(12,1)
    \psline[linecolor=red]{->}(7,2)(7.02,2.01)
    
    \rput(6.35,4.6){$x_{\star}$}
    
    \rput(5.45,.5){$\textcolor{red}{S_0}$}
    \rput(8.3,3.5){$\textcolor{red}{v_+}$}
    \rput(8.2,4.8){$\textcolor{red}{\ns}$}
    \rput(8.1,4.3){$\textcolor{red}{\theta}$}
    
    \rput(6.8,6.8){$\textcolor{red}{v_-}$}
    \psline[linecolor=red,linewidth=2pt]{->}(6,5)(6.4,7)
    
    \rput(1,1.75){Basin of $\bar{x}_1$}
    \rput(11,8.25){Basin of $\bar{x}_2$}
    
    \psline[linestyle=dotted](3,6.5)(2.5,4)(5.5,2.5)
    \rput(2.5,4){$\bullet$}
    \rput(2.2,4){$x$}
    \psline{<->}(6,5.1)(3,6.6)
    \rput(4.5,6.35){$\zeta_+(x)$}

    \psline[linecolor=blue,linewidth=2pt]{->}(0,4.6)(6,5)
    \psline[linecolor=blue,ArrowInside=->,ArrowInsideNo=1](6,5)(5,10)
    \psline[linecolor=blue,ArrowInside=->,ArrowInsideNo=1](6,5)(7,0)
    \pscurve[linecolor=blue]{->}(0,5)(4,7)(4.5,10)
    \pscurve[linecolor=blue]{->}(0,4)(4.5,3.1)(6.5,0)
    
    \rput(5,5.225){\textcolor{blue}{$v'_+$}}
    \rput(5.2,7.2){\textcolor{blue}{$v'_-$}}
    
    \psarc[linecolor=blue](6,5){1}{-175}{-100}
    \rput(5.1,4.1){\textcolor{blue}{$\gamma$}}

    \psarc[linecolor=blue](6,5){1}{80}{100}
    \rput(6,6.3){\textcolor{blue}{$\gamma'$}}

  \end{pspicture}
  \caption{The stable manifold $S_0 = x_{\star} + V_-$ and the unstable direction $v_+$ of the linearised relaxation dynamics are plotted with thick red lines. Thin red lines represent flow lines of the linearised dynamics $\dot{x}=\Ms(x-x_{\star})$. On this example, the coordinate $\zeta_+(x)$ is negative. The thick blue line is the incoming direction $v'_+$ of the instanton. Thin blue lines represent flow lines of the linearised fluctuation dynamics. The hatched area is the classically forbidden wedge, in which the linearised fluctuation dynamics does not describe fluctuation trajectories emanating from $\bar{x}_1$.}
  \label{fig:saddle:0}
\end{figure}
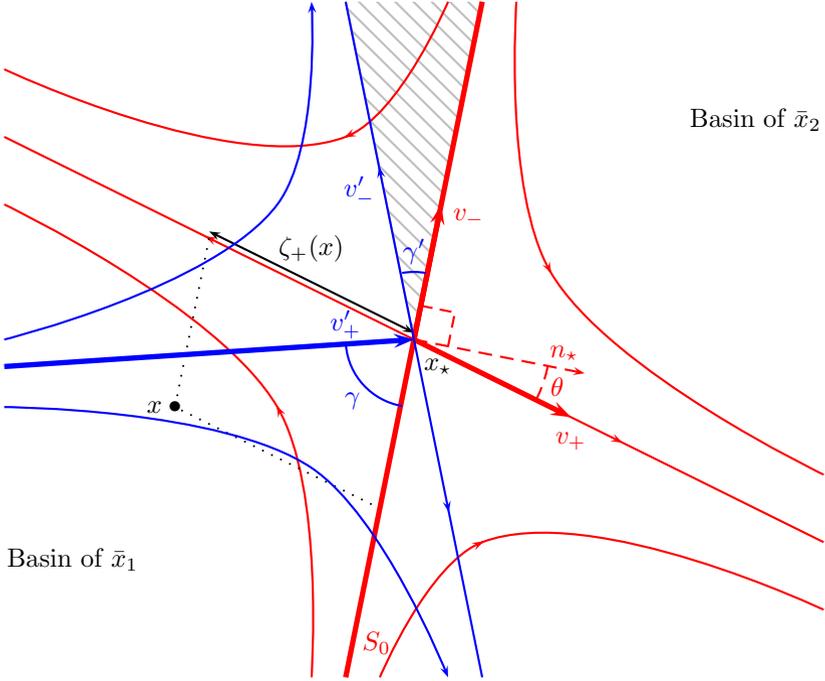

The linearisation of the fluctuation dynamics around the saddle-point writes
\begin{equation}
  \dot{x} = \Ns(x-x_{\star}), \qquad \Ns := \as\Hs + \Ds.
\end{equation}
The flow lines of this dynamics are plotted in blue on Figure~\ref{fig:saddle:0}. In order to provide a justification of this picture, let us detail explicit computations in dimension $d=2$, with $\as$ being the identity matrix. 

We first change the coordinates so that $x_{\star} = 0$ and 
\begin{equation}
  \Hs = \left(\begin{array}{cc}
    \mu_1 & 0\\
    0 & \mu_2
  \end{array}\right)
\end{equation}
with $\mu_1 < 0 < \mu_2$. Then~\eqref{eq:HDDtH} implies that there exists $\alpha \in \R$ such that
\begin{equation}
  \Ds = \left(\begin{array}{cc}
    0 & \alpha\\
    \alpha \rho & 0
  \end{array}\right)
\end{equation}
with $\rho := -\mu_1/\mu_2 > 0$. Let us denote $\mu := \mu_2 > 0$, so that $\mu_1 = -\rho\mu$. A direct computation shows that the eigenvalues of $\Ms$ are
\begin{equation}
  \begin{aligned}
    & \lambda^{\star}_+ := \frac{-\mu(1-\rho)+\mu(1+\rho)\sqrt{1+\frac{4\rho\alpha^2}{\mu^2(1+\rho)^2}}}{2} \geq \rho\mu > 0,\\
    & \lambda^{\star}_- := \frac{-\mu(1-\rho)-\mu(1+\rho)\sqrt{1+\frac{4\rho\alpha^2}{\mu^2(1+\rho)^2}}}{2} \leq -\mu < 0,
  \end{aligned}
\end{equation}
and we denote by $v_+$ and $v_-$ the respective associated eigenvectors. 

Clearly, the eigenvalues of $\Ns$ are $-\lambda^{\star}_+$ and $-\lambda^{\star}_-$, and we denote by $v'_+$ and $v'_-$ the respective associated eigenvectors. The direction of the instanton incoming at the saddle-point is given by the eigenvector $v'_+$ associated with the negative eigenvalue $-\lambda^{\star}_+$ of $\Ns$. A tedious but straightforward computation shows that the angle $\gamma$ between $v'_+$ and $v_-$ satisfies
\begin{equation}
  \sin(\gamma) = \frac{1}{\sqrt{1+(\alpha/\mu)^2}},
\end{equation}
which leads to the first conclusion that the assumption that $\Hs$ be nondegenerate prevents the incoming direction of the instanton at the saddle-point to be parallel to $S_0$. In contrast with Maier and Stein's results~\cite{MaiSte97}, this condition does not depend on the value of the ratio $\lambda^{\star}_+/|\lambda^{\star}_-|$, but rather on the sole assumption that $\mu>0$.

Likewise, the angle $\gamma'$ between $v_-$ and $v'_-$ vanishes if and only if $\alpha=0$, which leads to the second conclusion that if $\Ds \not=0$, then a region of $D_1$ does not contain any fluctuation trajectory emanating from $\bar{x}_1$: this is the {\em classically forbidden wedge} evidenced by Maier and Stein~\cite{MaiSte97}, which is hatched on Figure~\ref{fig:saddle:0}. In this area, the quasipotential $V(\bar{x}_1,\cdot)$ is not described by a transverse decomposition. Therefore the function $y \mapsto V(\bar{x}_1,y)$ is generically not smooth on the hyperplane $S_0$, which is the reason why the distribution of the point at which the Brownian particle reaches $S$ first is not Gaussian but rather skewed~\cite{MaiSte97}. More generally, it makes Kramers' analysis of the probability current across $S$ more delicate than the study presented in Section~\ref{s:bl}, which is the reason why we shall rather work on the hyperplane $S_{\eta}$ defined in~\eqref{eq:Seta} below.

\subsection{Sketch of the argument}\label{ss:sketch} 
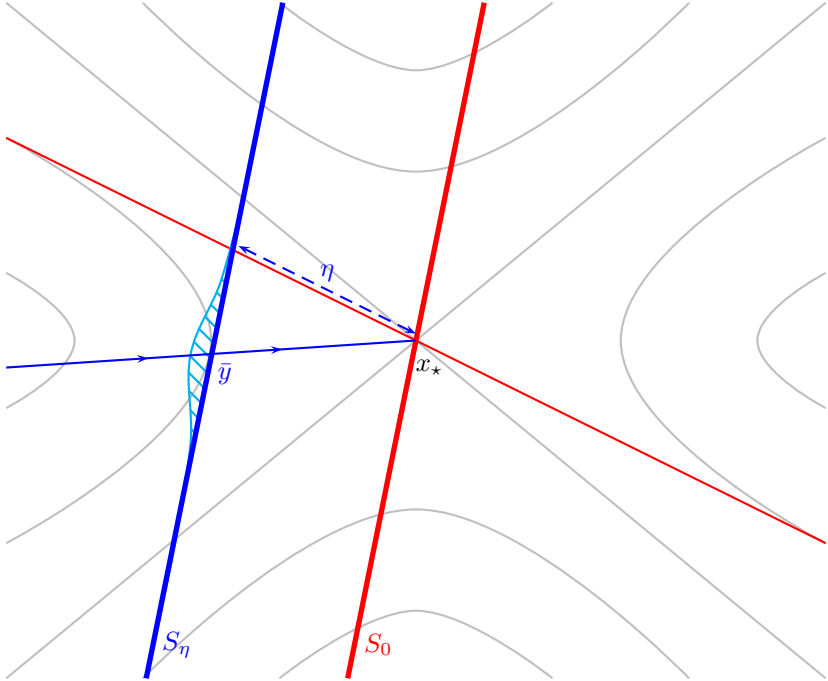
\begin{figure}[ht]
  \centering
  \psset{unit=.9cm}
  \begin{pspicture}(12,10)
    \psline[linecolor=lightgray]{-}(0,10)(12,0)
    \psline[linecolor=lightgray]{-}(0,0)(12,10)
    \pscurve[linecolor=lightgray]{-}(0,8)(3,5)(0,2)
    \pscurve[linecolor=lightgray]{-}(0,6)(1,5)(0,4)
    \pscurve[linecolor=lightgray]{-}(12,8)(9,5)(12,2)
    \pscurve[linecolor=lightgray]{-}(12,6)(11,5)(12,4)
    \pscurve[linecolor=lightgray]{-}(2,10)(6,7.5)(10,10)
    \pscurve[linecolor=lightgray]{-}(4,10)(6,9)(8,10)
    \pscurve[linecolor=lightgray]{-}(2,0)(6,2.5)(10,0)
    \pscurve[linecolor=lightgray]{-}(4,0)(6,1)(8,0)
    
    \psline[linecolor=red,linewidth=2pt]{-}(5,0)(7,10)
    \psline[linecolor=red]{-}(0,8)(12,2)

    \rput(6.2,4.6){$x_{\star}$}
    
    \rput(5.45,.5){$\textcolor{red}{S_0}$}
        
    \pscustom[linecolor=cyan,hatchcolor=cyan,fillstyle=vlines]{\pscurve(2.05,0)(2.7,3.5)(2.7,4.8)(3.15,5.9)(4.05,10)}

    \psline[linecolor=blue,linewidth=2pt]{-}(2.05,0)(4.05,10)
    \psline[linecolor=blue,linestyle=dashed]{<->}(6,5.1)(3.4,6.4)
    \rput(2.5,.5){$\textcolor{blue}{S_{\eta}}$}
    \rput(4.7,6){$\textcolor{blue}{\eta}$}
    \psline[linecolor=blue,ArrowInside=->,ArrowInsideNo=2](0,4.6)(6,5)
    \rput(3.2,4.5){$\textcolor{blue}{\bar{y}}$}
        
  \end{pspicture}
  \caption{The background gray lines are the equipotential lines in the neighbourhood of the saddle-point for the quadratic potential $\frac{1}{2} \scal{y-x_{\star}}{\Hs(y-x_{\star})}$. The hyperplane $S_{\eta}$ introduced in Subsection~\ref{ss:sketch} is plotted with the thick blue line. It is tangent to the equipotential line at the point $\bar{y}$, determined in~\S\ref{sss:minH}. The thin blue line is the instanton trajectory, it intersects $S_{\eta}$ at $\bar{y}$. The distribution of the current on $S_{\eta}$ is plotted in light blue, it is approximately Gaussian around $\bar{y}$ and has a standard deviation of the order of $\sqrt{\epsilon}$.}
  \label{fig:saddle}
\end{figure}

We are interested in the probability that, during the quasistationary phase in $D_1$, the particle crosses the separating hyperplane $S_0$ and drifts toward $\bar{x}_2$. Following the approach described in the introduction of this section, we decompose this event into two parts, and compute separately:
\begin{enumerate}[label=(\roman*), ref=\roman*]
  \item the probability that the particle first reaches the affine hyperplane 
  \begin{equation}\label{eq:Seta}
    S_{\eta} := x_{\star} - \eta v_+ + V_- = \{\zeta_+(x) = -\eta\},
  \end{equation}
  where $\eta \ll 1$ will be specified below (see Figure~\ref{fig:saddle}),
  \item the probability that a particle started from some point $y \in S_{\eta}$ actually crosses $S_0$ and leaves toward $\bar{x}_2$.
\end{enumerate}

We want to apply the results of Section~\ref{s:bl} in order to address the first point, and take as a domain $D_{\eta}$ the set $\{\zeta_+(x) < -\eta\}$. Although the assumption~\eqref{eq:asspartialD} of Section~\ref{s:bl} is not satisfied everywhere on $S_{\eta}$, it still holds in the neighbourhood of the point $\bar{y} \in S_{\eta}$ at which the instanton intersects $S_{\eta}$. This is sufficient for our analysis, since the probability current essentially concentrates around this point. Then using the assumption of smoothness of $V(\bar{x},y)$ in the neighbourhood of $\bar{y}$, we replace $V(\bar{x}_1,y)$ with its harmonic approximation
\begin{equation}
  V(\bar{x}_1,y) \simeq V(\bar{x}_1,x_{\star}) + \frac{1}{2} \scal{y-x_{\star}}{\Hs(y-x_{\star})},
\end{equation}
where we underline that the distance $|y-x_{\star}|$ is of order $\eta \ll 1$. We then deduce from~\eqref{eq:jsqt} that the probability that the particle crosses the hyperplane at some point $y \in S_{\eta}$ is given by
\begin{equation}\label{eq:jqst:temp}
  \begin{aligned}
    \scal{\jqst^{\epsilon}(y)}{\ns} & = \frac{\Cst(x_{\star})}{\epsilon^{d/2}} \exp\left(-\frac{V(\bar{x}_1,x_{\star})}{\epsilon} - \frac{\scal{y-x_{\star}}{\Hs(y-x_{\star})}}{2\epsilon}\right)\\
    & \quad \times \scal{\Ns(y-x_{\star})}{\ns}.
  \end{aligned}
\end{equation}
As a consequence, for $\sqrt{\epsilon} \ll \eta$, this distribution is approximately Gaussian, see Figure~\ref{fig:saddle}. The prefactor is computed in Subsection~\ref{ss:I} below.

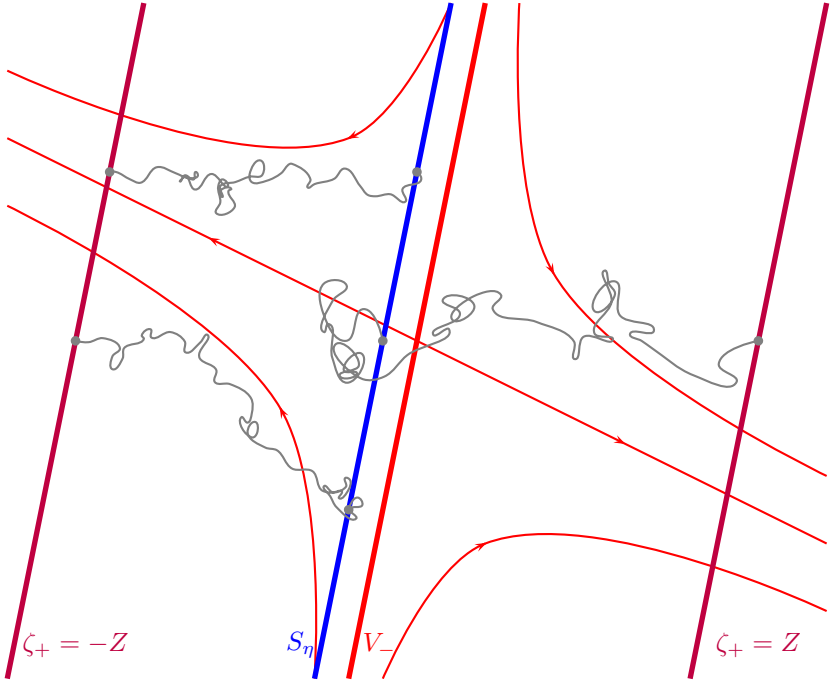
\begin{figure}[ht]
  \centering
  \psset{unit=.9cm}
  \begin{pspicture}(12,10)
    
    \psline[linecolor=red,linewidth=2pt]{-}(5,0)(7,10)
    \psline[linecolor=red,ArrowInside=->](6,5)(0,8)
    \psline[linecolor=red,ArrowInside=->](6,5)(12,2)
    
    \pscurve[linecolor=red](4.5,0)(4,4)(0,7)
    \psline[linecolor=red]{->}(4,4)(3.99,4.02)
    \pscurve[linecolor=red](7.5,10)(8,6)(12,3)
    \psline[linecolor=red]{->}(8,6)(8.01,5.98)
    
    \pscurve[linecolor=red](6.5,10)(5,8)(0,9)
    \psline[linecolor=red]{->}(5,8)(4.98,7.99)
    \pscurve[linecolor=red](5.5,0)(7,2)(12,1)
    \psline[linecolor=red]{->}(7,2)(7.02,2.01)
    
    
    \rput(5.45,.5){$\textcolor{red}{V_-}$}
%
%

    \psline[linecolor=blue,linewidth=2pt]{-}(4.5,0)(6.5,10)
    \rput(4.3,.5){$\textcolor{blue}{S_{\eta}}$}

    \psline[linecolor=purple,linewidth=2pt]{-}(0,0)(2,10)
    \psline[linecolor=purple,linewidth=2pt]{-}(10,0)(12,10)
    \rput(1,.5){\textcolor{purple}{$\zeta_+=-Z$}}
    \rput(11,.5){\textcolor{purple}{$\zeta_+=Z$}}
    
    \pscurve[linecolor=gray]{*-*}(5,2.5)(5.13,2.68)(5.2,2.57)(5.02,2.51)(5.11,2.37)(4.82,2.54)(4.73,2.7)(4.64,2.85)(4.76,2.76)(4.9,2.77)(4.86,2.67)(5,2.8)(5.02,2.94)(4.95,2.97)(4.96,2.84)(4.62,2.78)(4.47,3.07)(4.17,3.11)(3.95,3.38)(3.78,3.42)(3.52,3.64)(3.63,3.8)(3.66,3.64)(3.59,3.57)(3.43,3.87)(3.52,3.79)(3.58,3.92)(3.57,4.15)(3.26,4.36)(3.04,4.39)(3.06,4.57)(2.82,4.74)(2.81,4.8)(2.93,5.09)(2.83,4.98)(2.61,4.83)(2.6,4.99)(2.39,5.13)(2.11,5.05)(2.1,5.17)(1.97,5.08)(2,5.03)(1.71,4.89)(1.69,4.74)(1.54,4.58)(1.54,4.77)(1.65,4.89)(1.4,5.04)(1.19,4.98)(1,5)
    \pscurve[linecolor=gray]{*-*}(6,7.5)(6.06,7.38)(5.79,7.25)(5.83,7.06)(5.61,7.17)(5.39,7.14)(5.28,7.16)(5.17,7.29)(4.94,7.55)(4.79,7.4)(4.5,7.32)(4.4,7.27)(4.44,7.39)(4.17,7.62)(4.14,7.62)(4.01,7.48)(3.8,7.37)(3.58,7.35)(3.55,7.47)(3.71,7.67)(3.73,7.45)(3.55,7.43)(3.34,7.35)(3.12,7.29)(3.15,7.13)(3.13,7.23)(3.28,7.25)(3.24,7.22)(3.33,6.98)(3.06,6.89)(3.12,7.05)(3.09,7.29)(3.15,7.39)(3.23,7.29)(2.99,7.47)(2.75,7.24)(2.6,7.44)(2.53,7.36)(2.56,7.4)(2.6,7.45)(2.73,7.39)(2.76,7.5)(2.79,7.44)(2.65,7.41)(2.47,7.48)(2.19,7.57)(1.95,7.34)(1.8,7.39)(1.59,7.5)(1.5,7.5)
    \pscurve[linecolor=gray]{*-*}(5.5,5)(5.22,5.47)(4.96,5)(4.63,5.4)(4.64,5.85)(4.97,5.82)(4.77,5.6)(4.64,5.13)(4.7,5.25)(4.76,4.75)(4.98,4.38)(5.19,4.81)(4.93,4.56)(5.12,4.52)(4.87,4.48)(5.15,4.64)(4.81,4.95)(5.19,4.44)(5.52,4.47)(5.86,4.79)(6.47,5.17)(6.24,5.09)(6.63,5.4)(6.38,5.61)(6.84,5.52)(6.52,5.31)(6.83,5.66)(7.23,5.68)(7.11,5.47)(7.75,5.28)(8.3,4.99)(8.31,4.73)(8.44,5.06)(8.87,5.02)(8.6,5.11)(8.85,5.58)(8.75,6.04)(8.81,5.85)(8.9,5.69)(8.6,5.44)(8.9,5.88)(8.97,5.38)(9.47,5.15)(9.12,5.1)(9.29,5)(9.43,4.9)(10.05,4.49)(10.68,4.39)(10.42,4.66)(11,5)

  \end{pspicture}
  \caption{Typical and non-typical behaviour of the linearised diffusion process: when the particle starts on $S_{\eta}$, the drifts tends to bring it toward negative coordinates (as measured by $\zeta_+$), so that most trajectories hit $\{\zeta_+=-Z\}$ before hitting $\{\zeta_+=Z\}$. The commitor function describes the probability for a particle to perform the rare event of hitting $\{\zeta_+=Z\}$ before $\{\zeta_+=-Z\}$.}
  \label{fig:commitor}
\end{figure}

The second point is addressed by investigating the behaviour of the linearised diffusion process
\begin{equation}
  \dd \tilde{X}^{\epsilon}_t = \Ms(\tilde{X}^{\epsilon}_t-x_{\star}) \dd t + \sqrt{2\epsilon} \sigma(x_{\star}) \dd W_t
\end{equation}
far from the saddle-point. The particle following this linearised process started at a point of distance of order $\eta$ of $x_{\star}$ is more likely to drift away in the direction of $\bar{x}_1$ if its coordinate $\zeta_+(\tilde{X}^{\epsilon}_t)$ hits $-Z$ before hitting $Z$ for $Z \gg \eta$, see Figure~\ref{fig:commitor}. We therefore define the {\em commitor function} of the linearised dynamics by
\begin{equation}
  q^{\epsilon}(y) := \lim_{Z \to +\infty} \Pr[\tilde{\tau}^{\epsilon}_Z < \tilde{\tau}^{\epsilon}_{-Z}],
\end{equation}
where $\tilde{X}^{\epsilon}_0 = y$, and
\begin{equation}
  \tilde{\tau}^{\epsilon}_{\pm Z} := \inf\{t > 0: \zeta_+(\tilde{X}^{\epsilon}_t) = \pm Z\}.
\end{equation}
The commitor function is computed in Subsection~\ref{ss:commitor}.

Combining the definition of $q^{\epsilon}(y)$ with~\eqref{eq:jqst:temp}, we may finally conclude that the probability that the particle crosses the saddle-point during the quasistationary phase is given by
\begin{equation}\label{eq:lx1x2}
  \begin{aligned}
    & \lambda^{\epsilon}_{\bar{x}_1 \to \bar{x}_2} := \frac{\Cst(x_{\star})}{\epsilon^{d/2}} \exp\left(-\frac{V(\bar{x}_1,x_{\star})}{\epsilon}\right)\\
    & \times \int_{y \in S_{\eta}} q^{\epsilon}(y) \exp\left(- \frac{\scal{y-x_{\star}}{\Hs(y-x_{\star})}}{2\epsilon}\right) \scal{\Ns(y-x_{\star})}{\ns} \dd S(y),
  \end{aligned}
\end{equation}
for $\eta$ satisfying $\sqrt{\epsilon} \ll \eta \ll 1$. Then the average transition time is the inverse of this rate.


\subsection{Computation of the commitor function}\label{ss:commitor} Let us fix $y \in \R^d$ and $Z > 0$ with $Z \geq |\zeta_+(y)|$. For all $t \geq 0$, let us define
\begin{equation}
  z^{\epsilon}_t := \zeta_+(\tilde{X}^{\epsilon}_t),
\end{equation}
where $(\tilde{X}^{\epsilon}_t)_{t \geq 0}$ refers to the linearised dynamics started at $y$. Let us show that $(z^{\epsilon}_t)_{t \geq 0}$ solves a one-dimensional stochastic differential equation, so that $\Pr[\tilde{\tau}^{\epsilon}_Z < \tilde{\tau}^{\epsilon}_{-Z}]$ can be explicitly computed. On the one hand, using the decomposition of $\tilde{X}^{\epsilon}_t-x_{\star}$ on $v_+$ and $V_-$, we obtain
\begin{equation}
  \zeta_+(\Ms(\tilde{X}^{\epsilon}_t-x_{\star})) = \lambda^{\star}_+ z^{\epsilon}_t,
\end{equation}
while on the other hand,~\eqref{eq:zetap} shows that the process $(\zeta_+(\Ss W_t))_{t \geq 0}$ is a real-valued Brownian motion with variance $\scal{\as\ns}{\ns}/\cos^2\theta$. As a consequence, $(z^{\epsilon}_t)_{t \geq 0}$ is the (repulsive) Ornstein-Uhlenbeck process
\begin{equation}
  \dd z^{\epsilon}_t = \lambda^{\star}_+ z^{\epsilon}_t \dd t + \sqrt{2\epsilon}\frac{\sqrt{\scal{\as\ns}{\ns}}}{\cos\theta} \dd w_t,
\end{equation}
and for all $z \in [-Z,Z]$, the quantity 
\begin{equation}
  \bar{q}^{\epsilon}_Z(z) := \Pr[\tilde{\tau}^{\epsilon}_Z < \tilde{\tau}^{\epsilon}_{-Z} | z^{\epsilon}_0 = z]
\end{equation}
is obtained by solving the elliptic problem
\begin{equation}
  \frac{\epsilon\scal{\as\ns}{\ns}}{\cos^2\theta} \frac{\dd^2}{\dd z^2} \bar{q}^{\epsilon}_Z(z) + \lambda^{\star}_+ z \frac{\dd}{\dd z} \bar{q}^{\epsilon}_Z(z) = 0, \qquad z \in [-Z,Z],
\end{equation}
with the boundary conditions
\begin{equation}
  \bar{q}^{\epsilon}_Z(-Z) = 0, \qquad \bar{q}^{\epsilon}_Z(Z)=1.
\end{equation}
Integrating this equation yields
\begin{equation}
  \bar{q}^{\epsilon}_Z(z) = \frac{\displaystyle \int_{\xi = -Z}^z \exp\left(-\frac{\lambda^{\star}_+ \cos^2\theta}{2\epsilon\scal{\as\ns}{\ns}}\xi^2\right)\dd \xi}{\displaystyle \int_{\xi = -Z}^Z \exp\left(-\frac{\lambda^{\star}_+ \cos^2\theta}{2\epsilon\scal{\as\ns}{\ns}}\xi^2\right)\dd \xi},
\end{equation}
and taking the limit when $Z$ grows to infinity leads to the expression
\begin{equation}
  q^{\epsilon}(y) = \sqrt{\frac{\lambda^{\star}_+ \cos^2\theta}{2\pi\epsilon \scal{\as\ns}{\ns}}} \int_{\xi = -\infty}^{\zeta_+(y)} \exp\left(-\frac{\lambda^{\star}_+ \cos^2\theta}{2\epsilon\scal{\as\ns}{\ns}}\xi^2\right)\dd \xi
\end{equation}
of the commitor function.


\subsection{Integration over the hyperplane}\label{ss:I} The results of the previous subsection show that the commitor function $q^{\epsilon}(y)$ is uniform over $y \in S_{\eta}$, so that it can be taken out of the integral in the right-hand side of~\eqref{eq:lx1x2}. It therefore remains to compute
\begin{equation}
  I_{\eta}^{\epsilon} := \int_{y \in S_{\eta}} \exp\left(- \frac{\scal{y-x_{\star}}{\Hs(y-x_{\star})}}{2\epsilon}\right) \scal{\Ns(y-x_{\star})}{\ns} \dd S(y).
\end{equation}
We proceed in the following three steps. In~\S\ref{sss:minH}, we determine the point at which the exponential term assumes its maximum, and highlight the link with the instanton trajectory. In~\S\ref{sss:dethI} we express the value of $I_{\eta}^{\epsilon}$ in terms of the determinant of the restriction of the quadratic form $v \mapsto \scal{v}{\Hs v}$ to $V_-$. In~\S\ref{sss:dethH}, we connect this determinant with the determinant of $\Hs$ and the eigenvalue $\lambda^{\star}_+$. 

\subsubsection{Minimum of the potential along $S_{\eta}$}\label{sss:minH} We let $\bar{y}$ be the solution to the minimisation problem
\begin{equation}
  \min_{y \in S_{\eta}} \frac{1}{2}\scal{y-x_{\star}}{\Hs(y-x_{\star})},
\end{equation}
which rewrites  
\begin{equation}
  \min_{\scal{v_-}{\ns}=0} \frac{1}{2}\scal{-\eta v_+ + v_-}{\Hs(-\eta v_+ + v_-)},
\end{equation}
after the change of variable $y = x_{\star} -\eta v_+ + v_-$. Defining the Lagrangian
\begin{equation}
  \mathcal{L}(v_-, \lambda) := \frac{1}{2}\scal{-\eta v_+ + v_-}{\Hs(-\eta v_+ + v_-)} + \lambda \scal{v_-}{\ns},
\end{equation}
the optimality condition
\begin{equation}
  \nabla_{v_-} \mathcal{L} = 0
\end{equation}
implies that there exists $\lambda \in \R$ such that $\Hs (\bar{y}-x_{\star}) + \lambda \ns = 0$, that is to say
\begin{equation}
  \bar{y} = x_{\star} - \lambda \Hs^{-1}\ns.
\end{equation}

Let us prove that $\lambda = -\lambda^{\star}_+ \eta \cos\theta/\scal{\as\ns}{\ns}$, and that $\bar{y}-x_{\star}$ is an eigenvector of the matrix $\Ns$ for the eigenvalue $-\lambda^{\star}_+$. We first proceed to prove that
\begin{equation}\label{eq:NHn}
  \Ns \Hs^{-1} \ns = -\lambda^{\star}_+ \Hs^{-1} \ns.
\end{equation}
To this aim, we write
\begin{equation}
  \Ns \Hs^{-1} \ns = (\as\Hs + \Ds) \Hs^{-1} \ns = \as \ns - \Hs^{-1}\Ds^{\transp} \ns = - \Hs^{-1} \Ms^{\transp} \ns,
\end{equation}
where we have used~\eqref{eq:DHHDt} for the second equality. Now for all $u \in \R^d$, using the decomposition of $u$ on $v_+$ and $V_-$ yields
\begin{equation}
  \scal{\Ms^{\transp} \ns}{u} = \scal{\ns}{\lambda^{\star}_+ \zeta_+(u)v_+} = \lambda^{\star}_+ \scal{\ns}{u},
\end{equation}
thanks to~\eqref{eq:zetap}. As a consequence, 
\begin{equation}\label{eq:Mn}
  \Ms^{\transp} \ns = \lambda^{\star}_+\ns,
\end{equation}
which completes the proof of~\eqref{eq:NHn}. We now turn to the evaluation of $\lambda$, and first deduce from the series of equalities
\begin{equation}
  \scal{\bar{y}-x_{\star}}{\Hs(\bar{y}-x_{\star})} = -\lambda \scal{\bar{y}-x_{\star}}{\ns}
\end{equation}
and
\begin{equation}
  \begin{aligned}
    \scal{\bar{y}-x_{\star}}{\Hs(\bar{y}-x_{\star})} & = -\frac{1}{\lambda^{\star}_+}\scal{\Ns(\bar{y}-x_{\star})}{\Hs(\bar{y}-x_{\star})}\\
    & = -\frac{1}{\lambda^{\star}_+}\scal{\as\Hs(\bar{y}-x_{\star})}{\Hs(\bar{y}-x_{\star})}\\
    & = -\frac{\lambda^2}{\lambda^{\star}_+}\scal{\as\ns}{\ns}
  \end{aligned}
\end{equation}
that
\begin{equation}
  \lambda = \lambda^{\star}_+ \frac{\scal{\bar{y}-x_{\star}}{\ns}}{\scal{\as\ns}{\ns}}.
\end{equation}
But it follows from the definition of $S_{\eta}$ that $\scal{\bar{y}-x_{\star}}{\ns} = -\eta \cos\theta$, therefore we conclude that
\begin{equation}
  \lambda = - \frac{\lambda^{\star}_+\eta\cos\theta}{\scal{\as\ns}{\ns}}.
\end{equation}

As a conclusion, we have shown that the quadratic approximation of the potential assumes its minimum on $S_{\eta}$ at the point
\begin{equation}\label{eq:bary}
  \bar{y} = x_{\star} + \frac{\lambda^{\star}_+\eta\cos\theta}{\scal{\as\ns}{\ns}} \Hs^{-1}\ns,
\end{equation}
and that the vector $\bar{y}-x_{\star}$ satisfies
\begin{equation}\label{eq:yxvp}
  \Ns(\bar{y}-x_{\star}) = -\lambda^{\star}_+(\bar{y}-x_{\star}).
\end{equation}
Since $\Ns$ is the linearisation of the vector field driving the instanton dynamics, we are led to the following remarks.
\begin{enumerate}[label=(\roman*), ref=\roman*]
  \item The tangent hyperplanes to the equipotential hypersurfaces along the instanton trajectory are all parallel to each other (in the neighbourhood of the saddle point), and to the separating hyperplane between the basins of attraction of $\bar{x}_1$ and $\bar{x}_2$, see Figure~\ref{fig:saddle}.
  \item The eigenvalue of $\Ns$ associated with the direction of the instanton is the opposite of the eigenvalue of $\Ms$ associated with the unstable direction.
\end{enumerate}

\subsubsection{Gaussian approximation of $I^{\epsilon}_{\eta}$}\label{sss:dethI} Since $\bar{y}-x_{\star}$ is of order $\eta \gg \sqrt{\epsilon}$, we may use the approximation
\begin{equation}
  I_{\eta}^{\epsilon}\simeq \scal{\Ns(\bar{y}-x_{\star})}{\ns} \int_{y \in S_{\eta}} \exp\left(- \frac{\scal{y-x_{\star}}{\Hs(y-x_{\star})}}{2\epsilon}\right) \dd S(y),
\end{equation}
and~\eqref{eq:yxvp} yields
\begin{equation}
  \scal{\Ns(\bar{y}-x_{\star})}{\ns} = -\lambda^{\star}_+\scal{\bar{y}-x_{\star}}{\ns} = \lambda^{\star}_+ \eta \cos\theta.
\end{equation} 

Let us now compute the remaining Gaussian integral. The previous paragraph provides the canonical decomposition
\begin{equation}
  \scal{y-x_{\star}}{\Hs(y-x_{\star})} = \scal{y-\bar{y}}{\Hs(y-\bar{y})} + \scal{\bar{y}-x_{\star}}{\Hs(\bar{y}-x_{\star})}
\end{equation}
for all $y \in S_{\eta}$. On the one hand,
\begin{equation}
  \scal{\bar{y}-x_{\star}}{\Hs(\bar{y}-x_{\star})} = -\frac{\lambda^{\star}_+\eta^2\cos^2\theta}{\scal{\as\ns}{\ns}}.
\end{equation}
On the other hand, 
\begin{equation}
  \int_{y \in S_{\eta}} \exp\left(- \frac{\scal{y-\bar{y}}{\Hs(y-\bar{y})}}{2\epsilon}\right) \dd S(y) = \int_{v \in V_-} \exp\left(-\frac{\scal{v}{\Hs v}}{2\epsilon}\right) \dd S(y),
\end{equation}
so that denoting by $h \in \R^{(d-1) \times (d-1)}$ the matrix with coefficients $\scal{e_i}{\Hs e_j}$ where $(e_1, \ldots, e_{d-1})$ is an orthonormal basis of $V_-$, we obtain
\begin{equation}
  \int_{y \in S_{\eta}} \exp\left(- \frac{\scal{y-\bar{y}}{\Hs(y-\bar{y})}}{2\epsilon}\right) \dd S(y) = \sqrt{\frac{(2\pi\epsilon)^{d-1}}{\det h}}.
\end{equation}

We conclude that 
\begin{equation}
  I_{\eta}^{\epsilon} = \lambda^{\star}_+ \eta \cos\theta \sqrt{\frac{(2\pi\epsilon)^{d-1}}{\det h}} \exp\left(\frac{\lambda^{\star}_+\eta^2\cos^2\theta}{2\epsilon\scal{\as\ns}{\ns}}\right).
\end{equation}

\subsubsection{Expression of $\det h$}\label{sss:dethH} In this paragraph, we prove that $\det h$ is connected with $\lambda^{\star}_+$ and $\det \Hs$ through the identity
\begin{equation}\label{eq:dethH}
  \det \Hs = -\frac{\lambda^{\star}_+ \det h}{\scal{\as\ns}{\ns}}.
\end{equation}

In this purpose, we first establish a variant of the formula~\eqref{eq:DHHDt}. Using~\eqref{eq:HDDtH} and the definition of $\Ms$, we have
\begin{equation}
  \Hs\Ms + \Ms^{\transp}\Hs + 2\Hs \as \Hs = 0.
\end{equation}
Multiplying both sides by $\Hs^{-1}$, we deduce that $\Ms$ solves the stationary Lyapunov equation
\begin{equation}\label{eq:Lyap}
  \Ms \Hs^{-1} + \Hs^{-1} \Ms + 2 \as = 0.
\end{equation}

We now complete the orthonormal basis $(e_1, \ldots, e_{d-1})$ of $V_-$ introduced above by adding the normal vector $e_d := \ns$. In this basis, $\Hs$, $\Hs^{-1}$  and $\Ms$ assume the block decomposition
\begin{equation}
  \Hs = \left(\begin{array}{c|c}
    h & E^{\transp}\\
    \hline
    E & \gamma
  \end{array}\right), \quad \Hs^{-1} = \left(\begin{array}{c|c}
    * & F^{\transp}\\
    \hline
    F & \beta
  \end{array}\right), \quad \Ms = \left(\begin{array}{c|c}
    m & *\\
    \hline
    0 & \lambda^{\star}_+
  \end{array}\right).
\end{equation}
The $d-1$ first coefficients of the last line of $\Ms$ vanish because of the stability of $V_-$ by $\Ms$, while the fact that the last coefficient is $\lambda^{\star}_+$ immediately follows from the identity~\eqref{eq:Mn}. Evaluating~\eqref{eq:Lyap} with these block decompositions yields
\begin{equation}
  \lambda^{\star}_+\beta + \scal{\as\ns}{\ns} = 0.
\end{equation}
But by Cramer's rule, 
\begin{equation}
  \beta = \frac{1}{\det H} \det h,
\end{equation}
whence~\eqref{eq:dethH}.


\subsection{Conclusion} Introducing the function 
\begin{equation}
  \Phi(r) := \frac{1}{\sqrt{2\pi}} \int_{s=-\infty}^r \exp\left(-\frac{s^2}{2}\right) \dd s,
\end{equation}
so that, for all $y \in S_{\eta}$, the commitor function obtained Subsection~\ref{ss:commitor} writes
\begin{equation}
  q^{\epsilon}(y) = \Phi\left(-\eta\sqrt{\frac{\lambda^{\star}_+ \cos^2\theta}{\epsilon\scal{\as\ns}{\ns}}}\right),
\end{equation}
we compile the results of Subsection~\ref{ss:I} and obtain the following expression
\begin{equation}
  \begin{aligned}
    \lambda^{\epsilon}_{\bar{x}_1 \to \bar{x}_2} & = \frac{\Cst(x_{\star})}{\epsilon^{d/2}} \exp\left(-\frac{V(\bar{x}_1,x_{\star})}{\epsilon}\right) \Phi\left(-\eta\sqrt{\frac{\lambda^{\star}_+ \cos^2\theta}{\epsilon\scal{\as\ns}{\ns}}}\right)\\
    & \quad \times \lambda^{\star}_+ \eta \cos\theta \sqrt{\frac{(2\pi\epsilon)^{d-1}\lambda^{\star}_+}{|\det \Hs|\scal{\as\ns}{\ns}}} \exp\left(\frac{\lambda^{\star}_+\eta^2\cos^2\theta}{2\epsilon\scal{\as\ns}{\ns}}\right).
  \end{aligned}
\end{equation}
Using the asymptotic equivalence
\begin{equation}
  \Phi(r) \Sim_{r \to -\infty} \frac{1}{|r|\sqrt{2\pi}} \exp\left(-\frac{r^2}{2}\right),
\end{equation}
we observe that the terms depending on $\eta$ compensate and therefore lead to
\begin{equation}
  \lambda^{\epsilon}_{\bar{x}_1 \to \bar{x}_2} = \Cst(x_{\star}) \exp\left(-\frac{V(\bar{x}_1,x_{\star})}{\epsilon}\right) \lambda^{\star}_+ \sqrt{\frac{(2\pi)^{d-2}}{|\det \Hs|}}.
\end{equation}
Using the expression of $\Cst$ obtained in Section~\ref{s:stat} and noting that in the non-Gibbsianness correction, the instanton must be integrated for $t \in \R$, we conclude that the average transition time, given by the inverse of $\lambda^{\epsilon}_{\bar{x}_1 \to \bar{x}_2}$, has the expression
\begin{equation}
  \begin{aligned}
    \Exp[\tau^{\epsilon}_{\bar{x}_1 \to \bar{x}_2}] & \Sim_{\epsilon \dto 0} \frac{2\pi}{\lambda^{\star}_+} \sqrt{\frac{|\det H_{\star}|}{\det \Hess_x V(\bar{x}_1,\bar{x}_1)}} \exp\left(\int_{-\infty}^{+\infty} F(\rho_t)\dd t\right)\\
    & \quad \times \exp\left(\frac{V(\bar{x}_1,x)}{\epsilon}\right).
  \end{aligned}
\end{equation}
This formula is very similar to the classical Eyring-Kramers formula for reversible diffusions, with the quasipotential playing the role of the potential. The eigenvalue $\lambda^{\star}_+$ can be understood either as the unstable eigenvalue of the relaxation dynamics, or as the stable eigenvalue of the instanton dynamics around the saddle-point. Finally, the extra term involving the instanton trajectory is related with the prefactor of the quasistationary distribution and does not depend on the behaviour of the system around the saddle-point.



\begin{thebibliography}{1}

\bibitem{AriVdE07}
G.~Ariel and E.~Vanden-Eijnden.
\newblock Testing transition state theory on Kac-Zwanzig model.
\newblock {\em J. Stat. Phys.}, 126(1):43--73, 2007.

\bibitem{Arr89}
S. Arrhenius.
\newblock On the reaction velocity of the inversion of cane sugar by acids.
\newblock {\em J. Phys. Chem}, 4:226, 1889.

\bibitem{BaeKaf15}
Y. Baek and Y. Kafri.
\newblock Singularities in large deviation functions.
\newblock Preprint available at {\tt http://arxiv.org/abs/1505.05796}.

\bibitem{Bar15}
F. Barret.
\newblock Sharp asymptotics of metastable transition times for one dimensional
  {SPDE}s.
\newblock {\em Ann. Inst. Henri Poincar{\'e} Probab. Stat.}, 51(1):129--166,
  2015.

\bibitem{BarBovMel10}
F. Barret, A. Bovier, and S. M{{\'e}}l{{\'e}}ard.
\newblock Uniform estimates for metastable transition times in a coupled
  bistable system.
\newblock {\em Electron. J. Probab.}, 15(12):323--345, 2010.

\bibitem{Ber13}
N.~Berglund.
\newblock Kramers' law: validity, derivations and generalisations.
\newblock {\em Markov Process. Related Fields}, 19(3):459--490, 2013.

\bibitem{BerGen10}
N.~Berglund and B.~Gentz.
\newblock The {E}yring-{K}ramers law for potentials with nonquadratic saddles.
\newblock {\em Markov Process. Related Fields}, 16(3):549--598, 2010.

\bibitem{Ber14}
N. Berglund.
\newblock Noise-induced phase slips, log-periodic oscillations, and the Gumbel
  distribution.
\newblock Preprint available at {\tt http://arxiv.org/abs/1403.7393}.

\bibitem{BerDut13}
N. Berglund and S. Dutercq.
\newblock The Eyring-Kramers law for Markovian jump processes with symmetries.
\newblock {\em J. Theor. Probab.}, published online, 2015.

\bibitem{BerGen04}
N. Berglund and B. Gentz.
\newblock On the noise-induced passage through an unstable periodic orbit. {I}.
  {T}wo-level model.
\newblock {\em J. Stat. Phys.}, 114(5-6):1577--1618, 2004.

\bibitem{BerGen13}
N. Berglund and B. Gentz.
\newblock Sharp estimates for metastable lifetimes in parabolic {SPDE}s:
  {K}ramers' law and beyond.
\newblock {\em Electron. J. Probab.}, 18:(24)1--58, 2013.

\bibitem{BerGen14}
N. Berglund and B. Gentz.
\newblock On the noise-induced passage through an unstable periodic orbit {II}:
  {G}eneral case.
\newblock {\em SIAM J. Math. Anal.}, 46(1):310--352, 2014.

\bibitem{BouTou12}
F. Bouchet and H Touchette.
\newblock Non-classical large deviations for a noisy system with non-isolated
  attractors.
\newblock {\em J. Stat. Mech.},
  2012(05):P05028, 2012.

\bibitem{BovEckGayKle04}
A. Bovier, M. Eckhoff, V. Gayrard, and M. Klein.
\newblock Metastability in reversible diffusion processes I: Sharp asymptotics
  for capacities and exit times.
\newblock {\em J. Eur. Math. Soc.}, 6(4):399--424,
  2004.

\bibitem{CohLew67}
J.~K. Cohen and R.~M. Lewis.
\newblock A ray method for the asymptotic solution of the diffusion equation.
\newblock {\em IMA J. Appl. Math.}, 3(3):266--290, 1967.

\bibitem{DemZei10}
A. Dembo and O. Zeitouni.
\newblock {\em Large deviations techniques and applications}, volume~38 of {\em
  Stochastic Modelling and Applied Probability}.
\newblock Springer-Verlag, Berlin, 2010.
\newblock Corrected reprint of the second (1998) edition.

\bibitem{DykMilSme94}
M.~I. Dykman, M.~M. Millonas, and V.~N. Smelyanskiy.
\newblock Observable and hidden singular features of large fluctuations in
  nonequilibrium systems.
\newblock {\em Phys. Lett. A}, 195(1):53--58, 1994.

\bibitem{DykSmeMaiSil96}
M.I. Dykman, V.N. Smelyanskiy, R. S. Maier, and M. Silverstein.
\newblock Singular features of large fluctuations in oscillating chemical
  systems.
\newblock {\em J. Phys. Chem.}, 100(49):19197--19209, 1996.

\bibitem{Eyr35}
H. Eyring.
\newblock The activated complex in chemical reactions.
\newblock {\em J. Chem. Phys.}, 3(2):107--115, 1935.

\bibitem{FreWen12}
M.~I. Freidlin and A.~D. Wentzell.
\newblock {\em Random perturbations of dynamical systems}, volume 260.
\newblock Springer Science \& Business Media, 2012.

\bibitem{Gar85}
C.~W. Gardiner.
\newblock {\em Handbook of stochastic methods}, volume~13 of {\em Springer
  Series in Synergetics}.
\newblock Springer-Verlag, Berlin, second edition, 1985.
\newblock For physics, chemistry and the natural sciences.

\bibitem{DiGLeP15}
G. ~di Ges{\`u} and D.~Le Peutrec.
\newblock Small noise spectral gap asymptotics for a large system of nonlinear
  diffusions.
\newline\newblock Preprint available at {\tt http://arxiv.org/abs/1506.04434}.

\bibitem{Gra88}
R.~Graham.
\newblock Macroscopic potentials, bifurcations and noise in dissipative
  systems.
\newblock In {\em Noise in Nonlinear Dynamical Systems, Vol. 1}, pages
  225--278. Cambridge University Press, 1988.

\bibitem{GraHak71}
R.~Graham and H.~Haken.
\newblock Generalized thermodynamic potential for Markoff systems in detailed
  balance and far from thermal equilibrium.
\newblock {\em Zeit. Phys.}, 243(3):289--302, 1971.

\bibitem{GraTel84JSP}
R.~Graham and T.~T{\'e}l.
\newblock On the weak-noise limit of Fokker-Planck models.
\newblock {\em J. Stat. Phys.}, 35(5):729--748, 1984.

\bibitem{GraTel85}
R.~Graham and T.~T{\'e}l.
\newblock Weak-noise limit of Fokker-Planck models and nondifferentiable
  potentials for dissipative dynamical systems.
\newblock {\em Phys. Rev. A}, 31(2):1109, 1985.

\bibitem{HelKleNie04}
B. Helffer, M. Klein, and F. Nier.
\newblock Quantitative analysis of metastability in reversible diffusion
  processes via a Witten complex approach.
\newblock {\em Mat. Contemp.}, 26:41--86, 2004.

\bibitem{HwaHwaShe93}
C.-R. Hwang, S.-Y. Hwang-Ma, and S.-J. Sheu.
\newblock Accelerating {G}aussian diffusions.
\newblock {\em Ann. Appl. Probab.}, 3(3):897--913, 1993.

\bibitem{HwaHwaShe05}
C.-R. Hwang, S.-Y. Hwang-Ma, and S.-J. Sheu.
\newblock Accelerating diffusions.
\newblock {\em Ann. Appl. Probab.}, 15(2):1433--1444, 2005.

\bibitem{Kha11}
R. Khasminskii.
\newblock {\em Stochastic stability of differential equations}, volume~66.
\newblock Springer Science \& Business Media, 2011.

\bibitem{Kra40}
H.-A. Kramers.
\newblock Brownian motion in a field of force and the diffusion model of
  chemical reactions.
\newblock {\em Physica}, 7(4):284--304, 1940.

\bibitem{LanSwa61}
R. Landauer and J.A. Swanson.
\newblock Frequency factors in the thermally activated process.
\newblock {\em Phys. Rev.}, 121:1668, 1961.

\bibitem{Lan69}
J. S. Langer.
\newblock Statistical theory of the decay of metastable states.
\newblock {\em Ann. Phys.}, 54(2):258--275, 1969.

\bibitem{LelNiePav13}
T.~Leli{{\`e}}vre, F.~Nier, and G.~A. Pavliotis.
\newblock Optimal non-reversible linear drift for the convergence to
  equilibrium of a diffusion.
\newblock {\em J. Stat. Phys.}, 152(2):237--274, 2013.

\bibitem{LelRouSto10}
T. Leli{{\`e}}vre, M. Rousset, and G. Stoltz.
\newblock {\em Free energy computations}.
\newblock Imperial College Press, London, 2010.
\newblock A mathematical perspective.

\bibitem{Lud75}
D. Ludwig.
\newblock Persistence of dynamical systems under random perturbations.
\newblock {\em SIAM Rev.}, 17(4):605--640, 1975.

\bibitem{MaiSte93}
R.~S. Maier and D.~L. Stein.
\newblock Escape problem for irreversible systems.
\newblock {\em Phys. Rev. E}, 48(2):931, 1993.

\bibitem{MaiSte97}
R.~S. Maier and D.~L. Stein.
\newblock Limiting exit location distributions in the stochastic exit problem.
\newblock {\em SIAM J. Appl. Math.}, 57(3):752--790, 1997.

\bibitem{MatSch77}
B.~J. Matkowsky and Z. Schuss.
\newblock The exit problem for randomly perturbed dynamical systems.
\newblock {\em SIAM J. Appl. Math.}, 33(2):365--382, 1977.

\bibitem{MenSch14}
G. Menz and A. Schlichting
\newblock Poincar{\'e} and logarithmic Sobolev inequalities by decomposition of
  the energy landscape.
\newblock {\em Ann. Probab.}, 42(5):1809--1884, 2014.

\bibitem{Sch09}
Z. Schuss.
\newblock {\em Theory and applications of stochastic processes: an analytical
  approach}, volume 170.
\newblock Springer Science \& Business Media, 2009.

\bibitem{SchMat79}
Z. Schuss and B.~J. Matkowsky.
\newblock The exit problem: a new approach to diffusion across potential
  barriers.
\newblock {\em SIAM J. Appl. Math.}, 36(3):604--623, 1979.

\end{thebibliography}
\end{document}